\documentclass[journal,10pt]{IEEEtran}
\usepackage{amsmath,amssymb,amsfonts}
\usepackage{algorithm}
\usepackage{algorithmic}
\usepackage{graphicx}
\usepackage{textcomp}
\usepackage{xcolor}
\usepackage{pifont}
\usepackage{makecell}
\usepackage{markdown}
\usepackage{tcolorbox}
\usepackage{url}
\usepackage{subcaption}
\usepackage{url}
\usepackage{tikz}
\usepackage{siunitx}

\newcommand{\circled}[1]{\tikz[baseline=(char.base)]{\node[shape=circle,draw,inner sep=1pt] (char) {#1};}}

\title{LaMoSys3.5D: Enabling \underline{3.5D}-IC-Based \underline{La}rge Language \underline{Mo}del Inference Serving \underline{Sys}tems via Hardware/Software Co-Design}

\author{
    Qipan Wang~\IEEEmembership{Student Member,~IEEE},
    Zhe Zhang, Shuangchen Li, Hongzhong Zheng~\IEEEmembership{Member,~IEEE}, \\
    Zheng Liang~\IEEEmembership{Student Member,~IEEE},
    Yibo Lin~\IEEEmembership{Member,~IEEE},
    Runsheng Wang~\IEEEmembership{Member,~IEEE},
    Ru Huang~\IEEEmembership{Fellow,~IEEE}
    \thanks{This work was supported in part by the National Natural Science Foundation of China (Grant No. 62125401, 62034007), the Natural Science Foundation of Beijing, China (Grant No. Z230002), the Grant QYJS-2023-2303-B, the Beijing Outstanding Young Scientist Program (JWZQ20240101004), and the 111 Project (B18001).}
    \thanks{Q.~Wang is with the School of Advanced Interdisciplinary Studies and the School of Integrated Circuits, Peking University, Beijing 100871, China. Work done during an internship at Alibaba DAMO Academy.}
    \thanks{Z.~Liang is with the Department of Electrical Engineering and Computer Science, University of California at Berkeley, Berkeley, CA 94720 USA.}
    \thanks{Z.~Zhang, S.~Li and H.~Zheng are with the Alibaba DAMO Academy and Hupan Lab, China.}
    \thanks{Y.~Lin, R.~Wang, and R.~Huang are with the School of Integrated Circuits, Peking University, Beijing 100871, China; also with the Beijing Advanced Innovation Center for Integrated Circuits, Beijing 100871, China; and the Institute of Electronic Design Automation, Peking University, Wuxi 214135, China.}
    \thanks{Corresponding authors: Yibo Lin (\protect\url{yibolin@pku.edu.cn}) and Runsheng Wang (\protect\url{r.wang@pku.edu.cn}).}
}

\begin{document}

\maketitle

\begin{abstract}
The success of large language models (LLMs) amplifies the need for high-throughput, energy-efficient inference at scale. 3D-DRAM--based accelerators provide high memory bandwidth and therefore an opportunity to accelerate the bandwidth-bound decode phase. However, how to adequately balance compute density for prefill with bandwidth/capacity for decode remains open. Moreover, most prior designs do not target end-to-end serving, leaving the co-design of dataflow, parallel mapping, and scheduling underexplored.

To bridge the gap, we present \emph{LaMoSys3.5D}, to our knowledge the first scalable 3.5D-IC architecture for LLM serving. \emph{LaMoSys3.5D} composes heterogeneous 3D-DRAM chiplets on a 2.5D interposer: compute-rich chiplets for prefill and bandwidth-/capacity-rich chiplets for decode. 
To realize efficient serving, we adopt a hardware–software co-design spanning dataflow, parallel mapping, and introduce a thermal-aware modeling and hierarchical design-space exploration framework.
Across diverse LLMs and workloads, \emph{LaMoSys3.5D} improves throughput-per-watt over DGX-A100 systems by 62\% and achieves a $4.87\times$ better end-to-end latency (geo-mean) versus prior 3D designs. We further distill intriguing design guidelines for 3.5D-IC architectures and end-to-end inference serving.
\end{abstract}


\begin{figure}[t]
\centering
\includegraphics[width=1\linewidth]{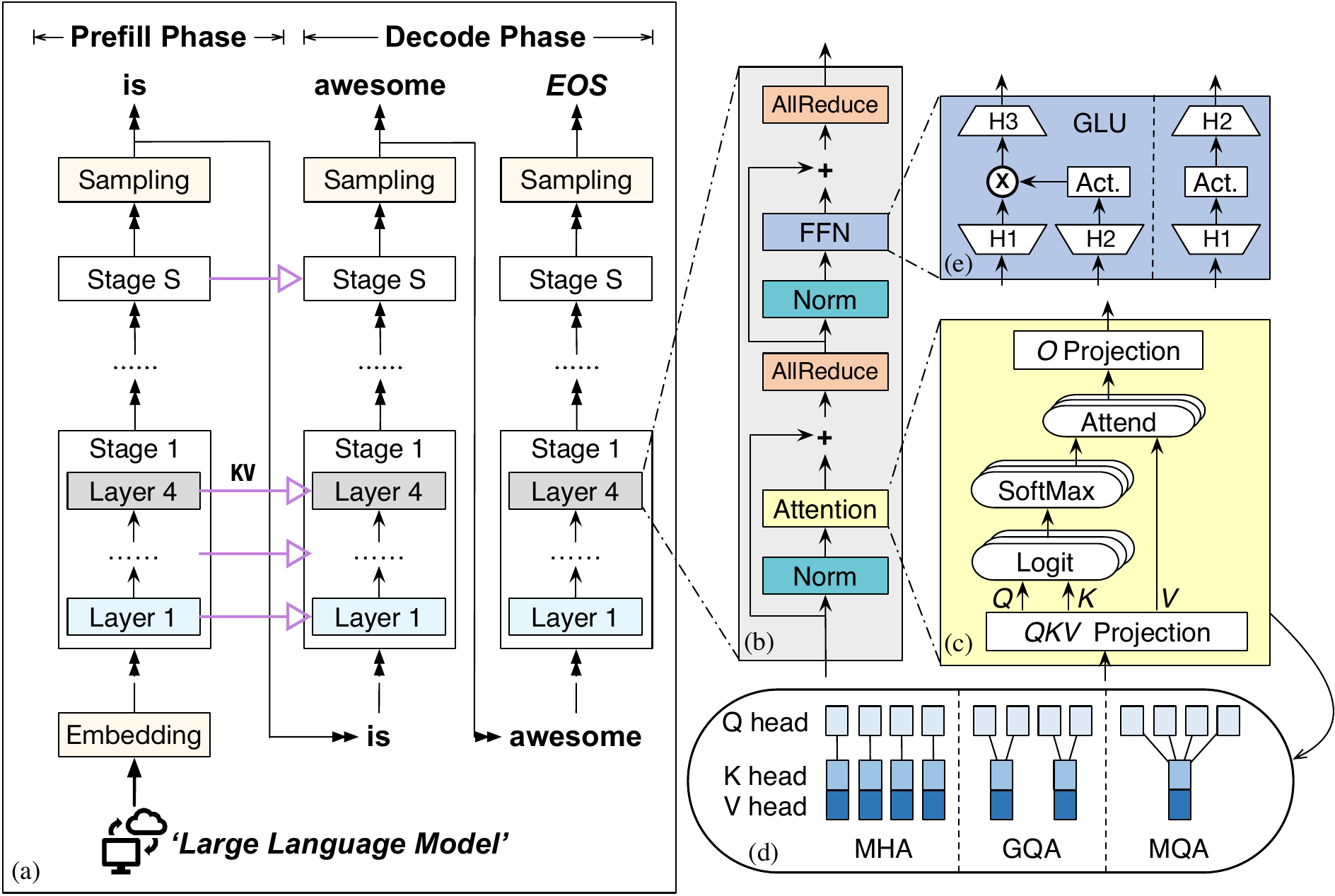}
\caption{LLM inference flow, including (a) the prefill and decode phases of the input sequence "Large Language Model", (b) architecture blocks of each transformer layer, (c) computation diagram of the self-attention block, (d) variants of attention operators, (e) variants of dense FFN operations.}
\label{fig:inference}
\end{figure}

\section{Introduction}

Large language models (LLMs) are permeating everyday applications~\cite{kaplan2020scaling}.
Their rapid adoption drives up inference cost and creates a pressing need for \emph{high-throughput, energy-efficient} serving at scale.
LLM inference has two phases with distinct characteristics (Fig.~\ref{fig:inference}(a)). \emph{Prefill} is the forward pass over the prompt and is compute-dominated. \emph{Decode} is an autoregressive loop that repeatedly accesses weights and the KV cache; it is bandwidth- and capacity-bound. Meeting these demands requires an inference hardware system that jointly optimizes compute and memory to sustain high throughput and energy-efficiency in both phases.

Recent reasoning-oriented models~\cite{plaat2024reasoning}—e.g., OpenAI o1~\cite{jaech2024openai}, DeepSeek-R1~\cite{guo2025deepseek}, and QwQ~\cite{QwenTeam2025}—prioritize decode speed, exposing limits of current hardware. HBM-based 2.5D ICs (e.g., GPU/TPU) dominate today, but HBM PHYs and routing consume substantial interposer area, constraining nearby logic. Some proposals offload decode to process-in-memory (PIM)-type architectures; however, this reduces compute density and thus harms prefill performance.
Hybrid-bonded (HB) 3D-DRAM offers an alternative: prototypes show higher memory bandwidth and lower energy/bit than HBM~\cite{fujun2020stacked,wang2023true}. 
Beyond a single 3D-DRAM chiplet, heterogeneous integration introduces a new degree of freedom: specialize {compute-rich} chiplets for prefill and {bandwidth-/capacity-rich} ones for decode.
This motivates a 3.5D-IC architecture that balances compute and memory for efficient LLM serving; however, realizing this vision raises three challenges.

\textbf{(1) Dataflow design to exploit 3D bandwidth.}
Classical dataflow for DDR/HBM-based systems minimizes off-chip traffic by maximizing on-chip SRAM reuse. To harness the high bandwidth of 3D{-}DRAM, TETRIS~\cite{gao2017tetris} reduces SRAM buffer size and pushes part of the computation toward memory; H2{-}LLM~\cite{2025h2llm} co-designs hardware and dataflow for a 3D{-}DRAM near-memory–processing (NMP) architecture; and 3D{-}TokSIM~\cite{20253dtoksim} couples 3D{-}DRAM with PIM and adopts a token-stationary schedule.
However, these designs largely depend on approximate analytical models or heuristic search.
Achieving an optimal dataflow for 3D-DRAM requires a complete specification of data reuse patterns and a scalable search over the whole mapping space.

\textbf{(2) Inference optimization on 3.5D-IC.}
Large-scale serving uses tensor and pipeline parallelism (TP/PP) and depends on efficient collective communications. Conventional GPU/TPU clusters adopt fully connected or ring topologies and high-speed fabrics (e.g., NVLink). Chiplet systems instead form side-by-side meshes over an interposer~\cite{CoWoS,mandalapu20243,shao2019simba}. Despite high die-to-die bandwidth, jointly optimizing the parallel mapping and scheduling to minimize intra-/inter-chiplet communication for diverse LLM workloads on such a topology remains unsolved.

\textbf{(3) Thermal risks in 3.5D-IC.}
The multilayer stack and compact chiplet layouts in 3.5D-IC increase heat density and impede heat removal, exacerbating thermal issues\cite{thermalimplication}. Elevated temperature degrades performance (e.g., refresh-induced bandwidth loss), increases leakage power, and undermines reliability~\cite{2024neurotap,10.1145/3419468}. 
Ignoring thermal effects during early design can lead to significant post-deployment inefficiencies. Therefore, thermal behavior should be modeled from the outset under realistic workloads, and its impact on performance and power should be quantified.

In response, we present \emph{LaMoSys3.5D}, to our knowledge \textbf{the first 3.5D-IC architecture designed for efficient LLM serving}. We assemble heterogeneous 3D-IC chiplets side-by-side on a 2.5D interposer, combining 3D vertical bandwidth/capacity with 2.5D lateral scalability. Our contributions are:
\begin{itemize}
\item We analyze LLM serving demands and introduce a scalable 3.5D-IC architecture tailored to modern, PD-disaggregated serving.
\item We define the complete dataflow–mapping space with 3D-DRAM, and propose a 3D-native $D^3$ dataflow that finds near-optimal mappings.
\item We develop novel parallel mapping strategies that reduce inter- and intra-chiplet communication cost.
\item We build an end-to-end, temperature-aware simulation framework for 3.5D-IC LLM inference serving, and use it to conduct hardware–software co-design and design-space exploration (DSE).
\end{itemize}

Across diverse LLMs and workloads, \emph{LaMoSys3.5D} improves throughput-per-watt over DGX\mbox{-}A100 systems by 62\% and achieves a $4.87\times$ better end-to-end latency (geo-mean) versus prior 3D designs. It also demonstrates high performance on workloads with long output sequences, 17.0$\times$ decode acceleration compared to A100, making it a promising candidate for future inference systems.
\section{Background}\label{sec:background}

\subsection{LLM Inference} \label{sec:metrics}
Modern {decoder-only} LLMs use the Transformer architecture~\cite{vaswani2017attention}, which stacks identical layers (Fig.\ref{fig:inference}(b)) of self-attention and feed-forward networks (FFNs)\cite{ramachandran2017swish}, implemented with either dense blocks\cite{achiam2023gpt,dubey2024llama} (Fig.\ref{fig:inference}(e)) or mixture-of-experts (MoE) modules\cite{liu2024deepseek}. 
During prefill, the model processes the input sequence in one forward pass to compute the distribution of the first output token; it also produces per-token key–value (KV) vectors that are stored as the {KV cache} for subsequent steps. 
During {decode} the model generates tokens autoregressively: at each step, it retrieves the KV cache, performs QKV projections, forms attention logits via $QK^\top$, applies \textsc{softmax}, attends to $V$, and performs FFN; the new K and V are appended to the cache.

We evaluate serving performance with four metrics: end-to-end latency (\textbf{E2E}, total time to produce the response), time to first token (\textbf{TTFT}, dominated by prefill), time between tokens (\textbf{TBT}, average per-token latency during decode), and throughput (\textbf{TPT}, tokens per second). 
Service level objects (SLOs) vary by task: batch workloads (e.g., summarization) emphasize throughput, while interactive workloads (e.g., dialog) prioritize \textbf{TTFT} and \textbf{TBT}. Reasoning models further tighten \textbf{TBT} due to inference-time scaling and longer sequences.

\subsection{LLM Serving Optimization}
Dynamic scheduling and parallelism are two common techniques that drive efficient LLM serving.
ORCA~\cite{yu2022orca} proposes {continuous batching}, which inserts new requests into a batch as soon as others finish, and {selective batching}, which batches only some operations while executing self-attention for requests in sequence.
The compute and memory demands of modern LLMs make multi-device inference attractive, especially when one device cannot hold the model and its KV cache. \emph{Pipeline parallelism} partitions layers into stages on distinct devices (Fig.\ref{fig:inference}(a)); overlapping requests across stages to improve throughput. \emph{Tensor parallelism} shards large tensor operations or attention heads across devices to accelerate, at the cost of inter-device communication and synchronization. For example, Megatron-LM\cite{shoeybi2019megatron} distributes parameters across model dimensions and requires two \emph{all-reduce} collectives per layer—after self-attention and after the FFN (Fig.~\ref{fig:inference}(b)). Logically, this is a \emph{reduce} followed by a \emph{multicast}.

\subsection{3.5D-IC Overview}
The term ``3.5D-IC''\cite{mandalapu20243} denotes packages that integrate one or more 3D-stacked dies within a 2.5D-IC system. Two packaging techniques enable this integration: hybrid bonding and the {silicon interposer}. 
HB provides ultra-fine-pitch die-to-die connections, often below \SI{10}{\micro\meter}, enabling thousands of links per \SI{}{\milli\meter\squared}. For example, stacked embedding DRAM (SeDRAM)\cite{fujun2020stacked,wang2023135,wang2023true} uses HB to stack DRAM onto logic, achieving bandwidth and energy efficiency that surpass those of GDDR and HBM. Similarly, AMD’s 3D V-Cache~\cite{agarwal20223d,smith202411} employs hybrid Cu–Cu bonding to stack an SRAM chiplet on a compute die. The 2.5D interposer connects chiplets, HBM, and passive components~\cite{naffziger2021pioneering} via TSVs and a back-end-of-line metal stack, linked with microbumps, enabling high-speed die-to-die communication across the package.

\begin{figure}[tbh]
    \centering
        \centering
        \includegraphics[width=.95\linewidth]{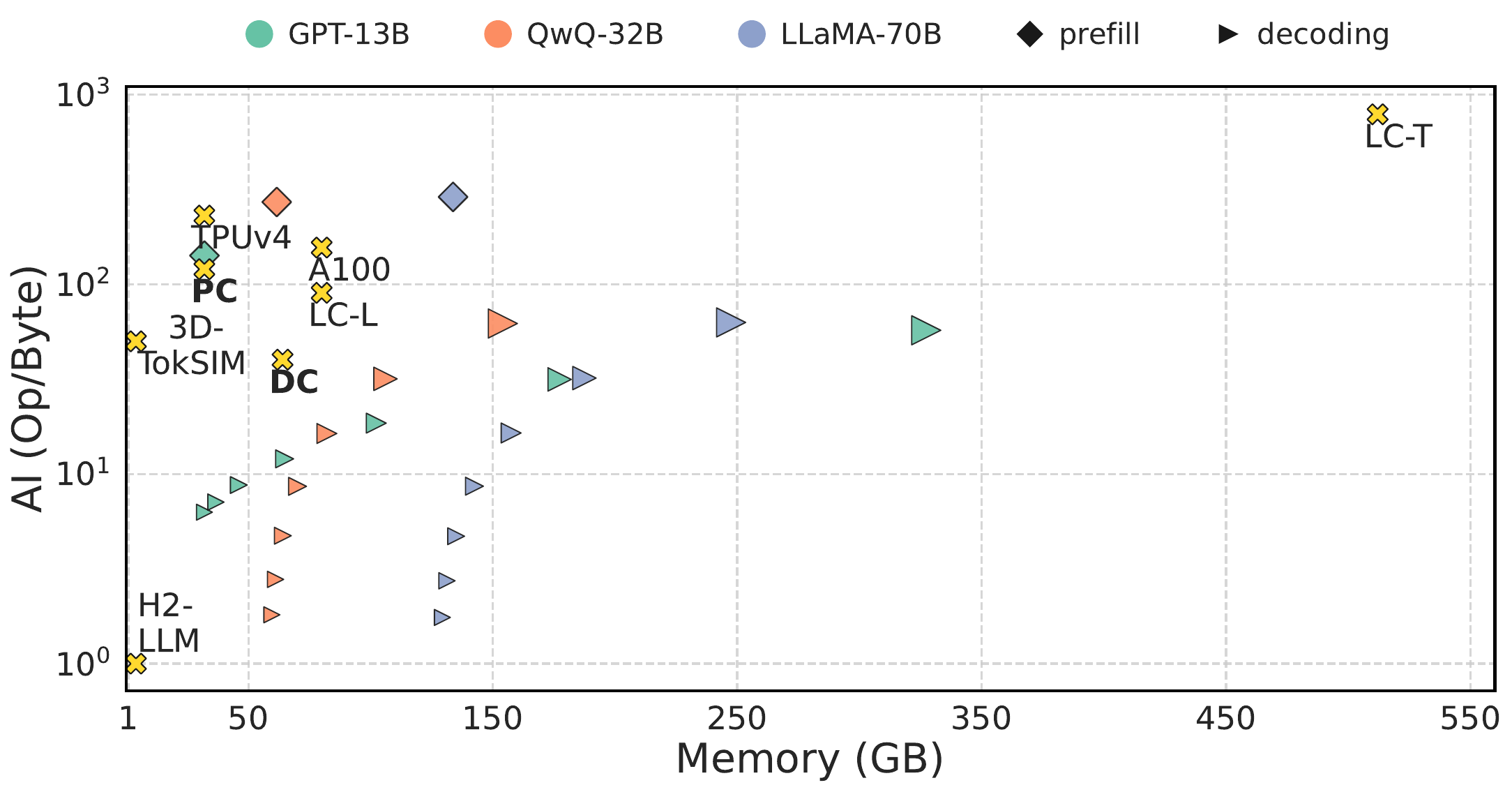}
    \centering
    \caption{AI and memory requirements for the prefill and decode phases of different LLMs. The batch size of decode ranges from 1 to 64, indicated by the size of the triangles.}
\label{fig:arithintens}
\vspace{-0.3cm}
\end{figure}

\begin{figure}[tbh]
  \centering
  \begin{minipage}[b]{0.46\linewidth}
    \centering
    \subfloat[]
    {\includegraphics[width=\linewidth]{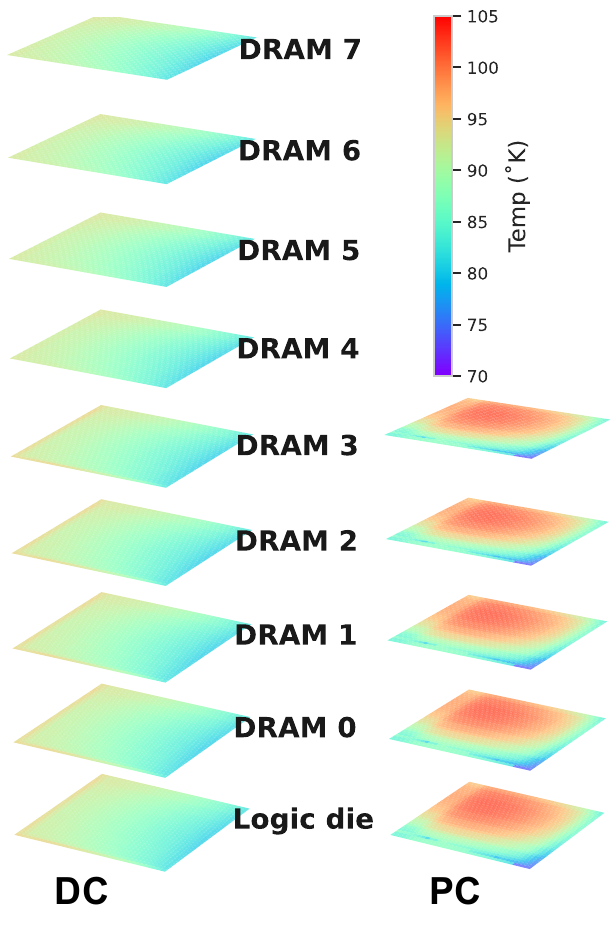}}
  \end{minipage}\hfill
  \begin{minipage}[b]{0.45\linewidth}
    \centering
    \subfloat[]
    {\includegraphics[width=\linewidth]{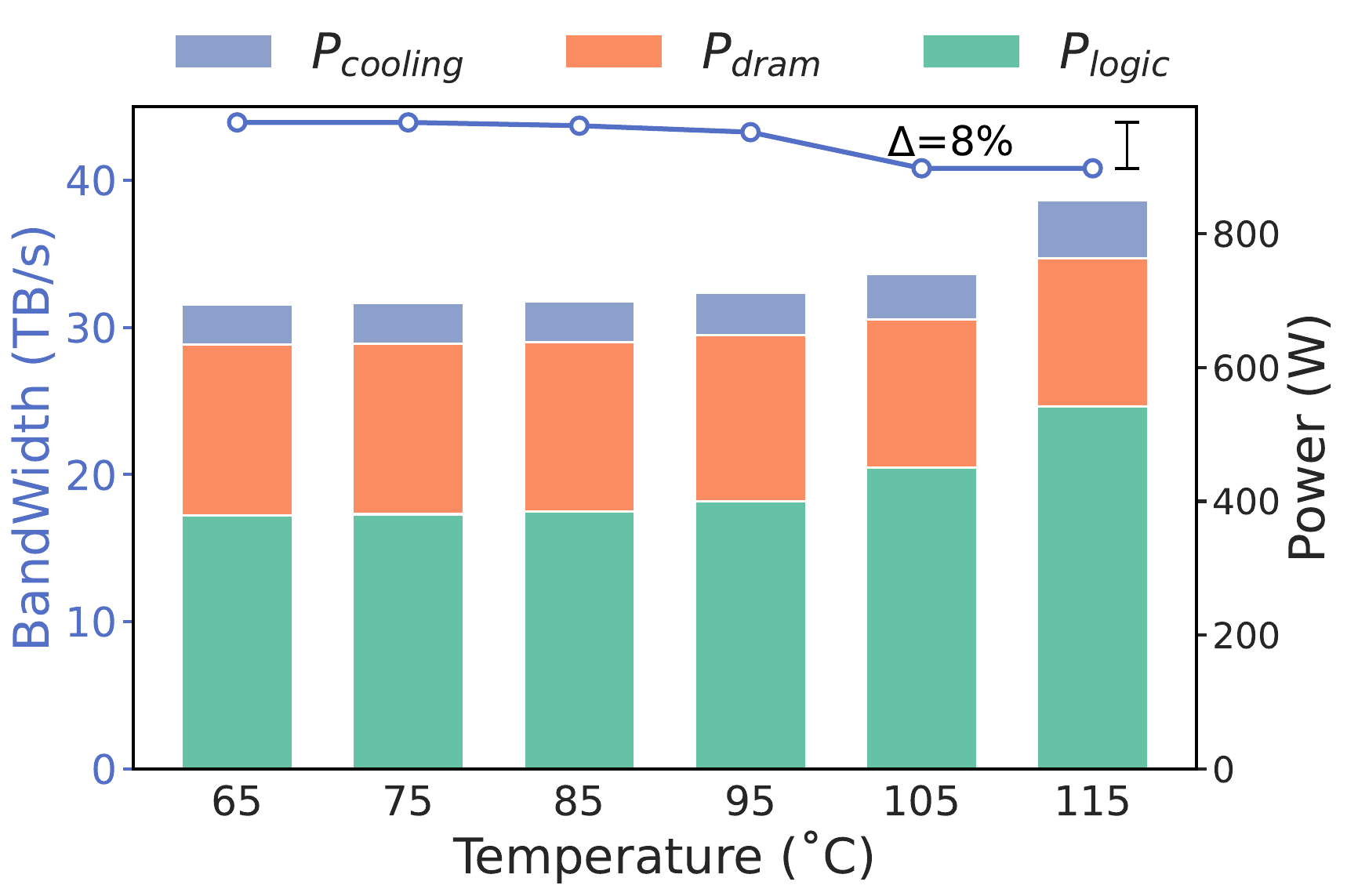}}\\
    \subfloat[]
    {\includegraphics[width=.96\linewidth]{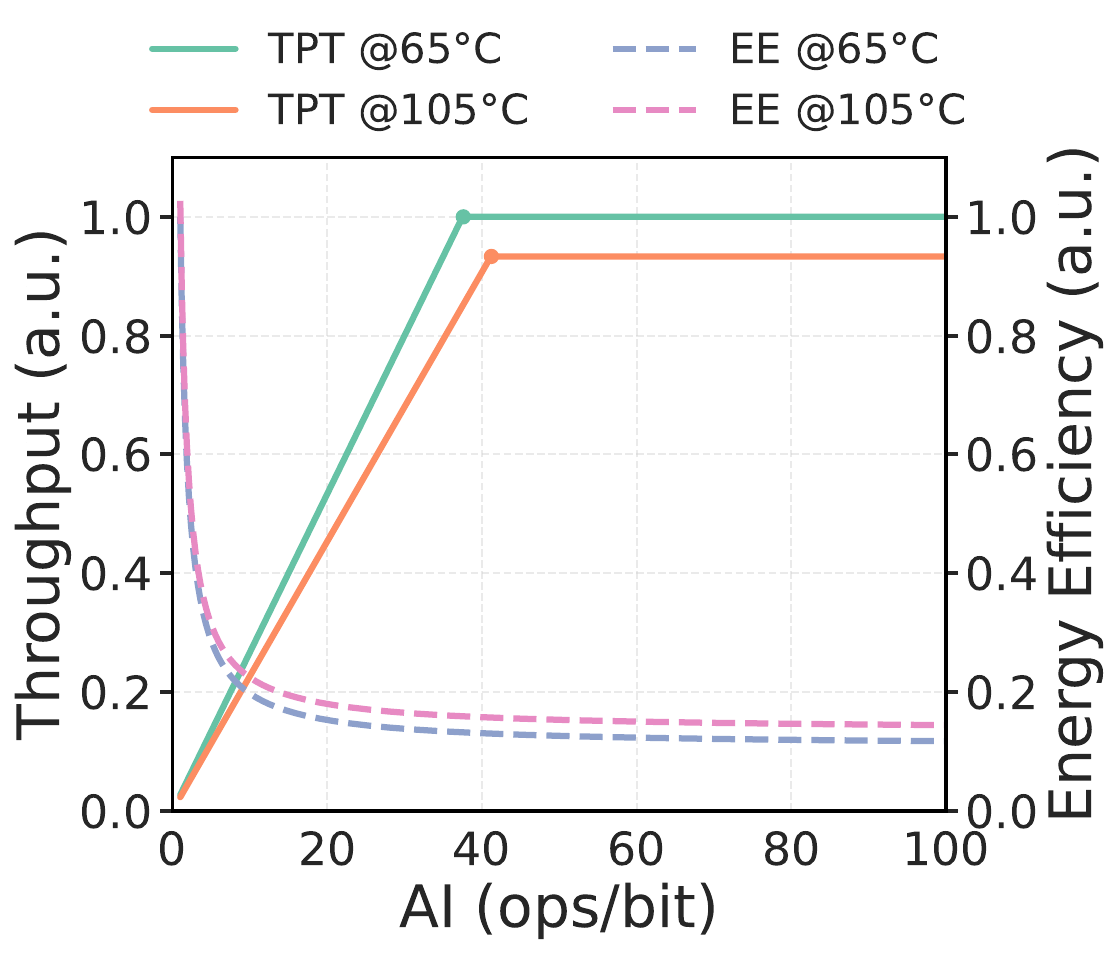}}
  \end{minipage}
  \caption{(a) Layer-wise 3D temperature map for our 3.5D-IC running certain workload. (b) Total power (7\,nm logic and 22\,nm DRAM) and chiplet bandwidth versus temperature. (c) Temperature-aware roofline model.}
  \label{fig:thermal-effects}
  \vspace{-0.3cm}
\end{figure}

\begin{table}[thb]
\centering
\caption{Comparison of Different 3D-DRAM Dataflow.}
\label{tab:dataflow_designs}
\resizebox{0.48\textwidth}{!}{
\begin{tabular}{|l|c|c|c|}
\hline
\textbf{Name} & \textbf{Compute Element} & \textbf{Data Reuse Pattern} & \textbf{Mapping Strategy} \\ \hline
TETRIS~\cite{gao2017tetris} & Mac Tree & Selective & Analytical \\ \hline
\cite{sharda2024accelerator} & Systolic Array & Input and Output & Heuristic \\ \hline
NicePIM~\cite{2024nicepim} & PIM & Input and Output & Heuristic \\ \hline
H2-LLM~\cite{2025h2llm} & Mac Tree & Selective & Heuristic \\ \hline
3D-TokSIM~\cite{20253dtoksim} & PIM Core & Input and Output & Analytical \\ \hline
$D^3$ (Ours) & Systolic Array & Selective & Exhaustive Search \\ \hline
\end{tabular}
}
\end{table}

\section{Motivation}\label{sec:motivation}

\subsection{Analysis of LLM Inference}
Prefill and decode stress hardware in different ways. Fig.\ref{fig:arithintens} reports arithmetic intensity (AI; FLOPs per byte moved) for representative LLMs—GPT3-13B\cite{achiam2023gpt}, QwQ-32B~\cite{bai2023qwen,yang2025qwen2}, and LLaMA3-70B~\cite{dubey2024llama}—across batch sizes 1–64 with input length 4096 and output length 2048. 
A roofline view shows a sharp phase asymmetry. Prefill has high AI and is compute-bound; decode has low AI and is bandwidth-/capacity-bound. Increasing batch size raises AI and improves utilization, but it also inflates DRAM capacity demand—especially for multi-head attention (MHA) models such as GPT-13B due to rapid KV-cache growth. In contrast, grouped-query attention (GQA)\cite{ainslie2023gqa}, used in LLaMA and QwQ, shrinks the cache footprint and eases bandwidth pressure (Fig.\ref{fig:inference}(d)).

These observations motivate {prefill–decode (PD) disaggregated serving}: map compute-intensive prefill to compute-centric devices and memory-bound decode to bandwidth-/capacity-centric devices. Prior work explores this idea either with heterogeneous pairings (e.g., H100–A100 in Splitwise~\cite{patel2024splitwise}, NPU–Flash in Cambricon-LLM~\cite{yu2024cambricon}, CPU–PIM in CENT~\cite{gu2025pim}, and others~\cite{zhu2025leveraging,kim2025ador}) or by partitioning homogeneous clusters into dedicated prefill and decode pools (e.g., DistServe, Sarathi-Serve~\cite{agrawal2024taming,hu2024inference}).

\vspace{-0.2cm}
\subsection{Opportunities with HB-based 3D-DRAM}
The rise of reasoning-oriented LLMs makes decode throughput the primary determinant of serving efficiency. We annotate arithmetic intensity (AI) and memory capacity for major platforms—A100, TPUv4, LLMCompass designs (\emph{LC-L}, \emph{LC-T})~\cite{zhang2024llmcompass}, 3D-TokSIM~\cite{20253dtoksim}, H2-LLM~\cite{2025h2llm}, and our 3D-IC chiplets (\emph{PC}, \emph{DC}). 
Systems with limited off-chip bandwidth— GDDR-based (e.g., \emph{LC-T}) or HBM-based (e.g., \emph{A100}, \emph{TPUv4}, \emph{LC-L})—exhibit a high AI, pushing most decode workloads into the bandwidth-bound regime. As a result, compute is underutilized and throughput is capped.

Recent efforts integrate HB-enabled 3D-DRAM with NPUs for transformers~\cite{2024nicepim,sharda2024accelerator,singh2024dram,2025h2llm,20253dtoksim}. As shown in Fig.~\ref{fig:arithintens}, these prototypes raise the memory roofline and ease bandwidth pressure (e.g., 3D-TokSIM, H2-LLM). Beyond bandwidth, HB integration reduces the interposer footprint compared to HBM, enabling higher on-die compute density and heterogeneous integration.

Motivated by this, we instantiate a 3D-DRAM chiplet template and, by configuring hardware parameters, derive {compute-rich} chiplets for prefill and {bandwidth-/capacity-rich} chiplets for decode. 
We assemble them side-by-side on a 2.5D interposer to form a 3.5D-IC, combining 3D vertical bandwidth/capacity with 2.5D lateral scalability.

\vspace{-0.2cm}
\subsection{Challenges Facing 3.5D-IC}

\paragraph{(1) SRAM-centric dataflow misfit 3D-DRAM}
Classical accelerator dataflow minimizes DRAM traffic by staging operands in SRAM to maximize reuse. For example, LLMCompass~\cite{zhang2024llmcompass} optimizes GEMM via exhaustive tiling and loop-order search under this SRAM-centric philosophy.
In contrast, emerging 3D-DRAM approaches SRAM in bandwidth and energy consumptions (e.g., 0.66–0.88~pJ/bit~\cite{fujun2020stacked,wang2023true}). In many kernels, the cost of fetching data directly from 3D-DRAM can rival buffering them in SRAM. 
Building on this observation, TETRIS~\cite{gao2017tetris} reduces on-chip SRAM size and pushes part of the computation toward memory; H2-LLM~\cite{2025h2llm} co-designs hardware and dataflow to exploit 3D-DRAM bandwidth with a prefill-aware dataflow; and 3D-TokSIM~\cite{20253dtoksim} combines 3D-DRAM with PIM and adopts a token-stationary flow that keeps token-related data in PIM while streaming weights. 
Tab.~\ref{tab:dataflow_designs} summarizes the compute element, data reuse pattern, and mapping strategies of representative works.
However, these designs rely on approximate analytical models (e.g., TETRIS, 3D-TokSIM) or heuristic algorithms (H2-LLM) to search the mapping, which risks missing the optimum. A 3D-native dataflow that flexibly allocates data across SRAM and 3D-DRAM and enables a scalable search for (near-)optimal mappings remains unsolved.

\paragraph{(2) Mesh-topology–aware parallel execution.}
CoWoS-class 2.5D technologies~\cite{CoWoS,mandalapu20243} provide scalable die-to-die bandwidth, yet efficient inter-/intra-chiplet communication for TP/PP remains open. TP/PP deployments commonly target {uniform-cost} topologies like rings or all-to-all. 
Our assembly instead forms a two-level mesh (intra- and inter-chiplet) with distance-dependent communication costs; standard ring all-reduce and related collectives become mismatched, inflating latency and KV traffic.
Prior art does not close this gap. {Gemini}\cite{cai2024gemini} maps CNN layers across chiplets in Simba-like systems using inter-/intra-layer partitioning. {PALM}\cite{fang2024palm} considers TP/PP mapping within two-level tiling structures for both training and inference. But these methods rely on heuristics without optimality guarantees and largely overlook KV-cache traffic.
These observations call for TP/PP mappings that best suit the two-level mesh.

\paragraph{(3) Thermal risks in 3.5D-ICs.} \label{sec:motiv-thermal}
The multi-layer stack configuration intensifies thermal risks~\cite{WANG2024126212}. 
Fig.~\ref{fig:thermal-effects}(a) shows a 3.5D-IC thermal map with peak temperatures above \SI{100}{\celsius}. When temperature exceeds \SI{85}{\celsius}, JEDEC mandates higher refresh (lower \emph{tREFI}), reducing effective DRAM bandwidth\cite{10.1145/3419468}. 
As shown in Fig.~\ref{fig:thermal-effects}(b), a \SI{40}{\celsius} rise (from \SI{65}{\celsius} to \SI{105}{\celsius}) increases logic leakage by $\sim$20\% and lowers peak DRAM power by $\sim$10\% due to refresh, cutting usable bandwidth by $\sim$10\% for our 3D-IC. 
The roofline therefore shifts unfavorably (Fig.~\ref{fig:thermal-effects}(c)), reducing throughput and worsening energy per operation.

Although prior work has addressed thermal-aware design, cooling, and runtime management for 2.5D/3D ICs~\cite{2024wangatplace,salvi2021review,2024neurotap}, {early-stage} thermal evaluation for 3.5D-IC is still lacking, posing performance and reliability risks. 
This gap motivates workload-aware, transient thermal analysis integrated into early architecture design. 

To address these issues, we first propose a 3D-native dataflow, then devise mesh-aware parallel mapping and scheduling strategies, and build a unified full-stack framework coupling hardware modeling, mapping optimization, and performance and thermal analysis. On this basis, we perform hierarchical, thermal-aware DSE across hardware (chiplet and system) and software (dataflow and parallel mapping).
\begin{figure}[tbh]
    \centering
    \includegraphics[width=1\linewidth]{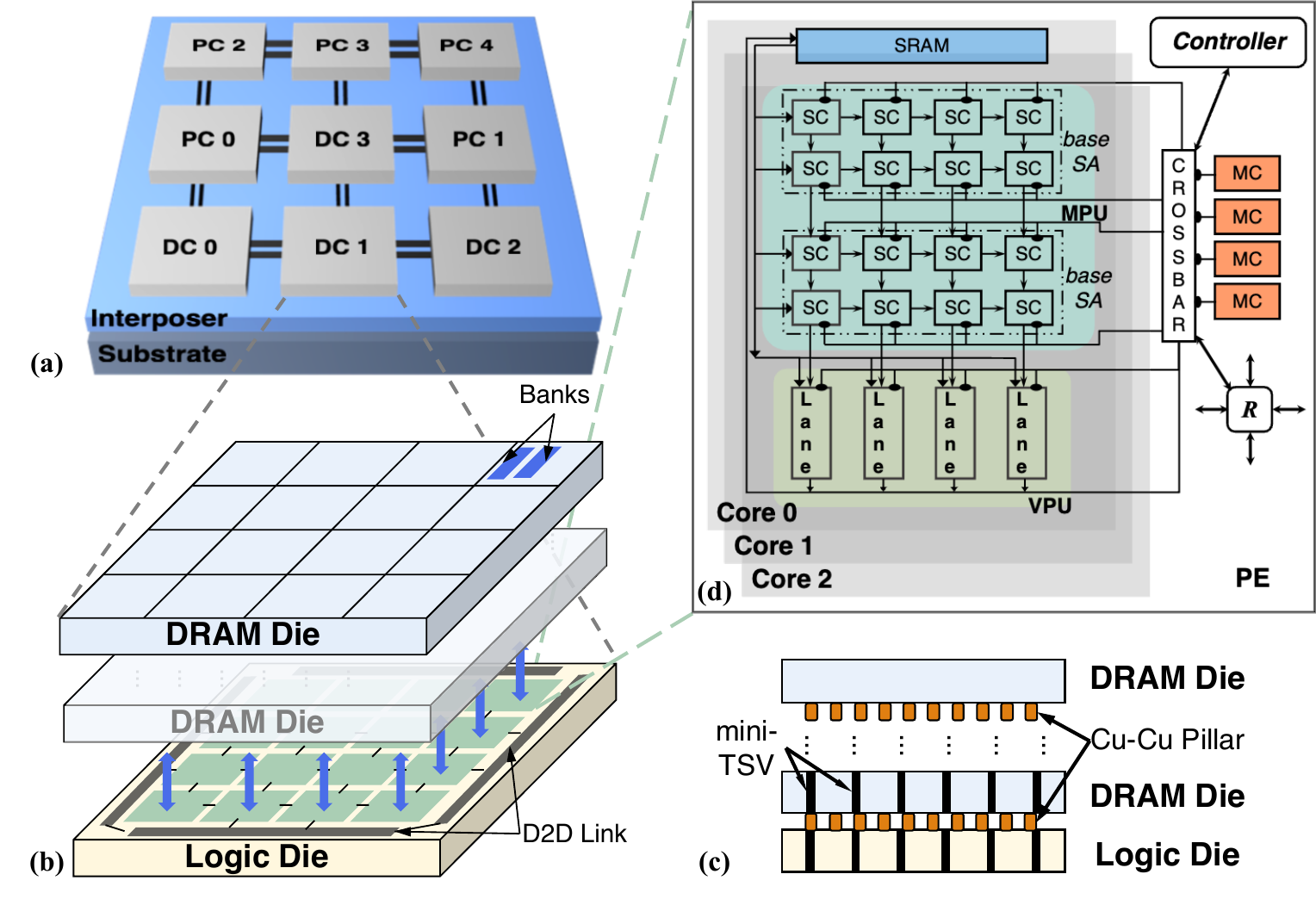}
    \caption{Hardware template of the proposed \emph{LaMoSys3.5D}.}
\label{fig:template}
\vspace{-0.2cm}
\end{figure}

\section{3.5D-IC Architecture}\label{sec:hardware}

\subsection{Overall Design}
As shown in Fig.~\ref{fig:template}(a), multiple chiplets are arranged in a mesh~\cite{shao2019simba} and interconnected via die-to-die (D2D) links. Each chiplet is implemented as a DRAM-on-logic 3D-IC: DRAM layers are stacked directly above the logic die using HB as shown by Fig.~\ref{fig:template}(b) and (c). The memory and logic dies may use different technology nodes; in all cases, the logic-die area is bounded by the DRAM die footprint, and total power is constrained by thermal headroom.

At the system level, we provision chiplets with similar functionality but different compute--memory ratios to compose scalable serving clusters. Prefill chiplets (PC) are compute-leaning and require fewer memory banks/layers due to modest bandwidth and capacity demands. Decode chiplets (DC) are memory-leaning: they increase the number of banks/layers and allocate more area to memory controllers (MCs) while reducing compute. Distinct parallelism requirements in prefill and decode~\cite{zhong2024distserve} are met by tailoring the number of PEs per chiplet. PC and DC variants share a unified core design and DRAM IP, enabling a PD-disaggregated serving system when clustered together. 
At the chiplet boundary, D2D links on all four edges connect to a network-on-package (NoP) over a passive interposer, following SIAM-style hierarchy~\cite{2021SIAM}; we use AIB~2.0~\cite{tang2023arvon} as the PHY, provisioned for up to \SI{1}{\tera\byte/\second} aggregate bandwidth.

\subsection{3D-DRAM Design}\label{sec:dram}
We adopt a fine-grained die-stacked DRAM organization~\cite{wang2023true,chen2024high}. The stack comprises $N_{\text{layer}}$ DRAM dies, each implementing $N_{\text{bank}}$ banks. Every bank exposes an $N_{\text{io}}$-bit TSV data interface, creating independent vertical channels that exploit bank-level parallelism. TSV fabric traverses the stack and terminates at MCs on the logic die via HB, enabling very wide interfaces and few-ns read/write latency in advanced nodes.
DRAM efficiency is shaped by access scheduling and refresh policy. We employ first-come first-served (FCFS) scheduling with a closed-page policy~\cite{li2018performance}. A layer-interleaved strategy~\cite{zhu20133d} staggers bank operations across layers to overlap activation and column access. Because capacitors leak, periodic refresh is required: the canonical retention window is \SI{64}{ms} below \SI{85}{\celsius} and halves for every \SI{10}{\celsius} increase. We use distributed all-bank refresh~\cite{JESD79-4,thakkar2016massed}, scheduling refresh commands across banks/layers within the retention interval to minimize visible bandwidth loss.

\subsection{Processing Element (PE) Architecture}
As shown in Fig.~\ref{fig:template}(d), each PE integrates a five-port router, a high-radix crossbar, a RISC-V controller, $N_{\text{core}}$ compute cores, and $N_{\text{bank}}$ MCs. The router coordinates data movement among PEs and the D2D endpoints; flit width is configurable to meet bandwidth targets. The crossbar provides low-latency connectivity among intra-PE endpoints (cores, MCs, controller). Every DRAM channel terminates at a dedicated MC, enabling independent, parallel accesses. Within each core, a matrix processing unit (MPU) is paired with a lightweight vector processing unit (VPU) that handles common elementwise/normalization kernels (e.g., activations, softmax, LayerNorm/RMSNorm) and participates in reductions, similar to Ascend’s vector unit~\cite{liao2019ascend}. This consolidation keeps the PE description succinct while preserving the functionality required to sustain the MPU.

\subsection{MPU Design}\label{sec:mpu}
Conventional systolic arrays (SAs) with $SA_{\text{rows}}\!\times\!SA_{\text{cols}}$ cells deliver high throughput for large-batch \textsc{GEMM} by pipelining operand streams~\cite{kim2023full}. However, utilization collapses for small-batch \textsc{GEMM}—and especially for \textsc{GEMV} in decode (e.g., per-head attention)—because when the effective matrix height $M<SA_{\text{rows}}$, only a fraction $M/SA_{\text{rows}}$ of rows is active.
Elegant prior efforts alleviate this via heterogeneous datapaths (e.g., coupling SAs with MAC trees~\cite{kim2025ador}) or reconfigurable sub-arrays like Planaria~\cite{9251939} and RSA~\cite{tang2022hardware}. 
While effective within their target regimes, these designs misalign with our 3D-DRAM design: (i) they underutilize abundant vertical bandwidth from DRAM channels, (ii) they introduce fine-grained control overheads that bloat area/latency. Consequently, we pursue a simpler, bandwidth-aware alternative.

To this end, we partition a large systolic array (SA) into multiple {short--wide} \textbf{baseSAs} (fewer rows, same columns), as shown in Fig.~\ref{fig:template}(c).
This organization preserves high-throughput \textsc{GEMM} while better matching \textsc{GEMV}: additional baseSAs raise utilization when $M$ is small or $K$ is modest (e.g., 128 per head). Projection and logit layers—dominant decode bottlenecks—thereby see substantial speedups that scale with the number of simultaneously active baseSAs (e.g., input batch size).
Each \emph{baseSA} can fetch operands from SRAM or stream them directly from 3D-DRAM through the crossbar; control logic scales along the row dimension, keeping area overhead low. This design enables selective SRAM bypass and DRAM streaming and aligns with our dataflow.
\begin{figure}[tbh]
    \centering
    \includegraphics[width=1\linewidth]{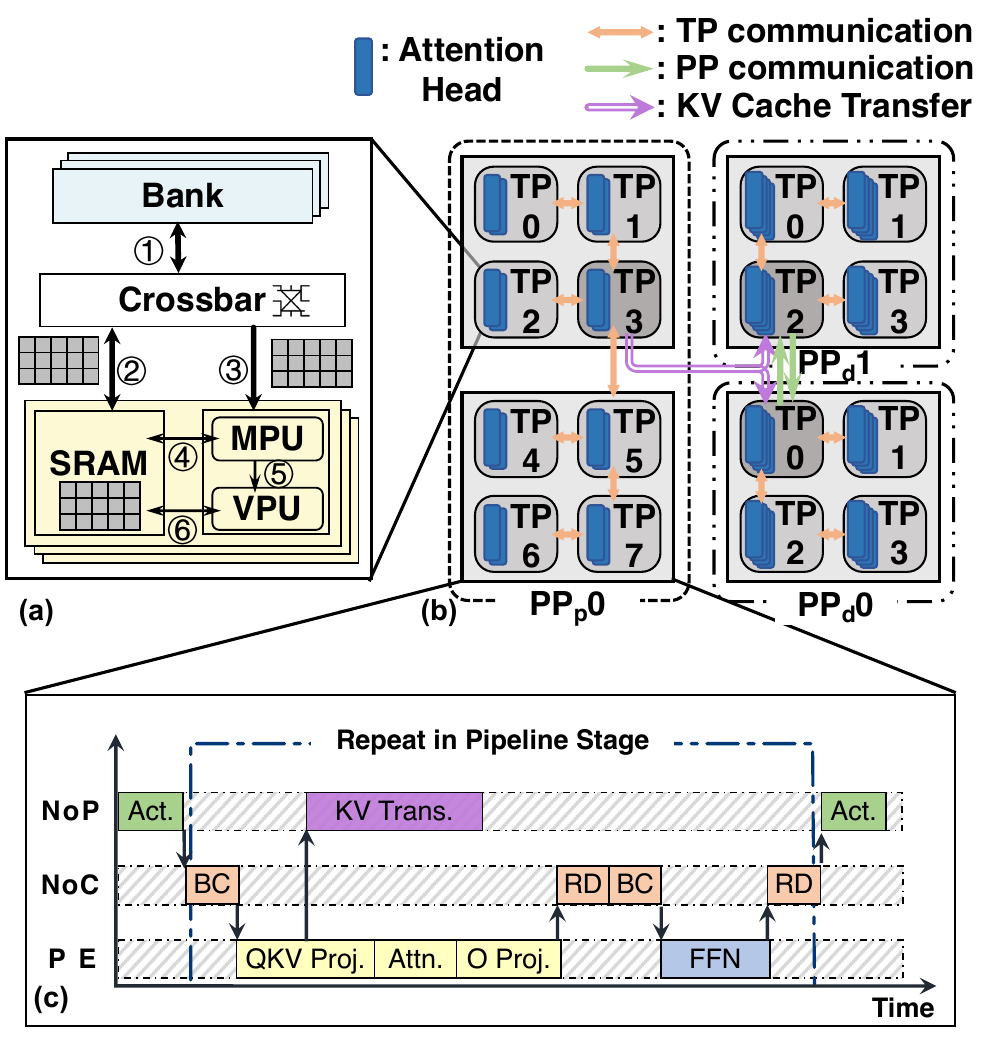}
    \caption{(a) Intra-PE data path and executions; (b) representative PP/TP mapping scheme with (TP, PP) = (8, 1) and (4, 2) for prefill and decode, respectively; (c) Timing of operations of one stage executed on \emph{PC}.}
\label{fig:dataflow}
\vspace{-0.2cm}
\end{figure}

\begin{algorithm}[tbh]
\caption{\textsc{D$^3$} Intra-PE Dataflow Search}
\label{alg:dataflow}
\begin{algorithmic}[1]
\REQUIRE Tensor sizes $M,N,K$, SRAM budget $S_{\text{buf}}$, cost model $\mathcal{C}$
\STATE $\mathcal{L}_{\min}\!\leftarrow\!\infty$, $(T_M^*,T_N^*,T_K^*,RU^*)\!\leftarrow\!\bot$
\STATE Enumerate candidate tiles $\mathcal{T}_M,\mathcal{T}_N,\mathcal{T}_K$
\FOR{$(T_M,T_N,T_K)\in \mathcal{T}_M \times \mathcal{T}_N \times \mathcal{T}_K$}
    \STATE $
    \mathcal{R} \leftarrow
    \begin{aligned}[t]
       \{\text{IRU}\mid T_MT_K \le S_{\text{buf}}\}
       &\cup \{\text{WRU}\mid T_NT_K \le S_{\text{buf}}\}\ \cup \\
       \{\text{ORU}\mid T_MT_N \le S_{\text{buf}}\}
       &\cup \{\text{ARU}\mid \sum_{\text{cyc}} T_M T_N \le S_{\text{buf}}\}
    \end{aligned}$
    \FOR{$\text{RU}\in\mathcal{R}$}
        \STATE $\ell \leftarrow \mathcal{C}(T_M,T_N,T_K,\text{RU})$
        \IF{$\ell<\mathcal{L}_{\min}$} \STATE $\mathcal{L}_{\min}\!\leftarrow\!\ell$; $(T_M^*,T_N^*,T_K^*,RU^*)\!\leftarrow\!(T_M,T_N,T_K,\text{RU})$ \ENDIF
    \ENDFOR
\ENDFOR
\RETURN $(T_M^*,T_N^*,T_K^*,RU^*)$
\end{algorithmic}
\end{algorithm}

\begin{figure*}[tbh]
    \centering
    \includegraphics[width=0.9\linewidth]{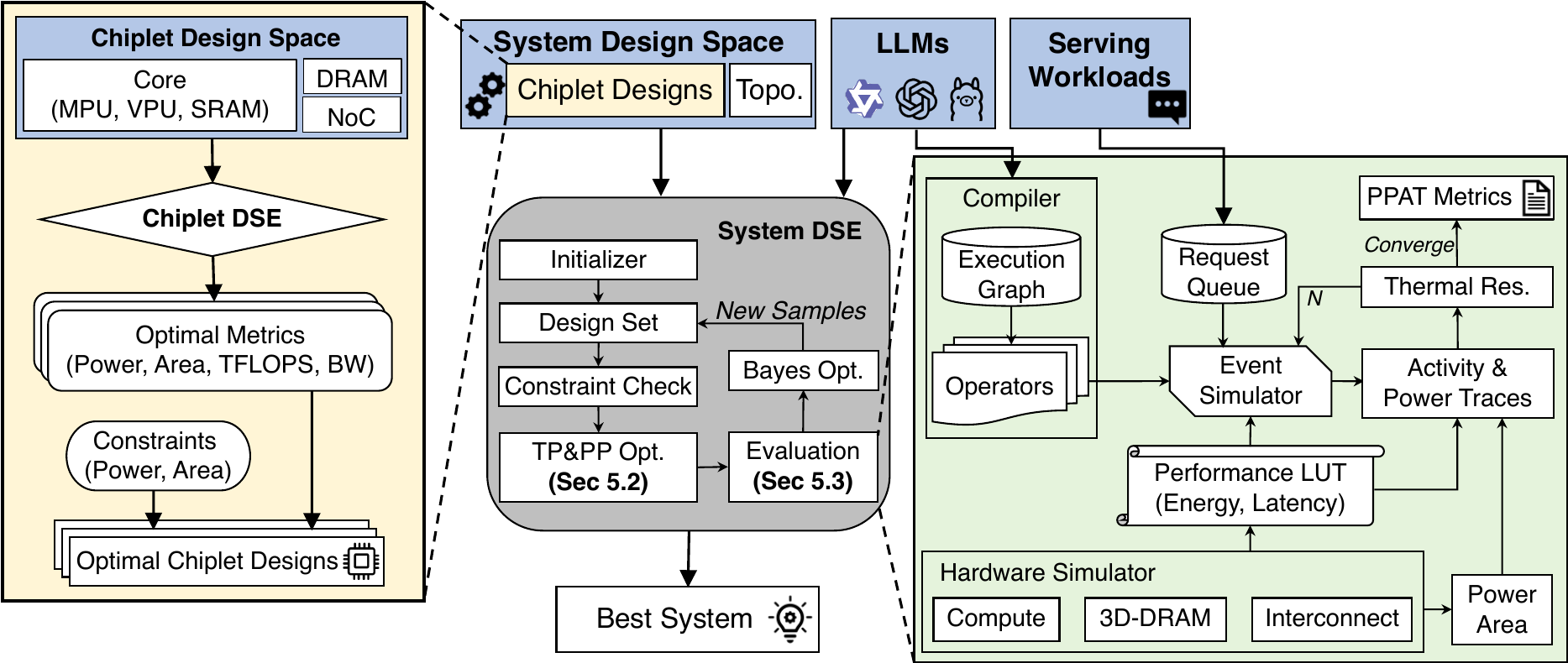}
    \caption{Overview of our simulation and design framework.}
\label{fig:framework}
\vspace{-0.2cm}
\end{figure*}

\section{Software Framework}\label{sec:framework}
This section presents our software framework. We begin with the \emph{D$^3$} (direct-dram-delivery) dataflow design intra PE (Sec.~\ref{sec:dataflow}), followed by the parallel mapping and scheduling scheme for serving (Sec.~\ref{sec:serving}). 
Then, we describe our performance simulator (Sec.~\ref{sec:simulator}) and end with DSE methodology (Sec.~\ref{sec:dse}).

\subsection{Intra-PE Dataflow Design} \label{sec:dataflow}
For a general \textsc{GEMM} $C \leftarrow A\!\times\!B + C$ with $A\!\in\!\mathbb{R}^{M\times K}$, $B\!\in\!\mathbb{R}^{K\times N}$, $C\!\in\!\mathbb{R}^{M\times N}$, all operands are tiled along $(M,N,K)$ during calculation. 
Prior SRAM-centric dataflow typically stage certain tiles in SRAM (\circled{1} $\rightarrow$ \circled{2}) to maximize operand reuse, with multiply–accumulate operations performed on the MPU and reductions in the VPU (\circled{4} $\rightarrow$ \circled{5} $\rightarrow$ \circled{6}) before writing back to DRAM, with the operations (\circled{1}-\circled{6}) defined in Fig.~\ref{fig:dataflow}(a). 
While our {D$^3$} dataflow determines flexibly whether an operand tile is {temporarily staged in SRAM for reuse} (\circled{1}$\rightarrow$\circled{2}$\rightarrow$\circled{4}) or {streamed directly from the stacked DRAM} (\circled{1}$\rightarrow$\circled{3}) to MPU. 
This enables direct-DRAM delivery for selected tiles (e.g., model weights) while staging only important tiles for reuse, reducing SRAM pressure without sacrificing throughput.
We classify reuse policies as: 
(i) \textbf{IRU} (Input-Reuse): stage $A$-tiles; 
(ii) \textbf{WRU} (Weight-Reuse): stage $B$-tiles;
(iii) \textbf{ORU} (Output-Reuse): stage $C$-tiles/partial outputs; 
(iv) \textbf{ARU} (All-Reuse): stage $(A,B,C)$ tiles simultaneously.
A reuse policy is feasible if its SRAM footprint fits $S_{\text{buf}}$ (e.g., IRU requires $T_M T_K\!\le\!S_{\text{buf}}$). 
We employ exhaustive exploration rather than the analytical formulations as ~\cite{gao2017tetris,20253dtoksim} to identify the globally optimal mapping. The process is shown in Alg.~\ref{alg:dataflow}: (i) enumerate candidate tile sizes; (ii) for each tiling, find feasible reuse policies $\mathcal{R}$ that satisfy the SRAM capacity constraint; (iii) simulate the latency of each pair $(\text{tiling},\text{RU})$ with the simulator in \S\ref{sec:core-sim}; and (iv) select the best mapping $(T_M^*,T_N^*,T_K^*,RU^*)$.

\begin{algorithm}[t]
\caption{TP Grouping via MILP}
\label{alg:ilp_cluster}
\begin{algorithmic}[1]
\REQUIRE PE coordinates $\mathcal{P}=\{(x_i,y_i)\}_{i=1}^{P}$; group size $TP$; weight $w_{\text{inter}}$ for inter-group costs
\ENSURE Groups $\mathcal{G}=\{G_1,\dots,G_K\}$ of size $TP$ and per-group centers $(c^x_k,c^y_k)$
\STATE $K \gets \left\lfloor P/TP \right\rfloor$ \COMMENT{max \#stages (groups)}
\STATE \textbf{Decision variables:}
\STATE \quad $c_{i,k}\!\in\!\{0,1\}$ (PE $i$ assigned to group $k$);\;
$m^k_x,M^k_x,m^k_y,M^k_y\in\mathbb{Z}$ (group $k$ bounding box);\;
\STATE \textbf{Constraints:}
\STATE \quad \emph{Assignment:} $\sum_{k=1}^{K} c_{i,k}\le 1,\ \forall i$;\quad $\sum_{i=1}^{P} c_{i,k}=TP,\ \forall k$
\STATE \quad \emph{Bounding box (x-dimension; y analogous).} Let $B_x \!\gets\! (\max_i x_i-\min_i x_i)+1$, \qquad $M^k_x \ge x_i - B_x(1-c_{i,k}),$ \qquad $m^k_x \le x_i + B_x(1-c_{i,k}),\ \forall i,k$
\STATE \quad \emph{Centers}
$c^x_k = \tfrac{1}{2}(M^k_x+m^k_x),\ \forall k$
\STATE \textbf{Objective:} minimize
$J = \sum_k D_k+w_{\text{inter}}\cdot \sum_{\mathcal{A}} \textsc{Dist}(k,k\!+\!1)$,
For adjacent pairs $\mathcal{A}=\{(k,k\!+\!1)\,|\,k=1..K\!-\!1\}$ 
\STATE \quad \emph{Span upper bound (proxy for all-reduce cost):}
$D_k \ge (M^k_x-m^k_x) + (M^k_y-m^k_y),\ \forall k$
\STATE \quad \emph{Inter-group center distances.} $\textsc{Dist}(k,k\!+\!1) \gets \|c^x_{k}-c^x_{k+1}\|+\|c^y_{k}-c^y_{k+1}\|$
\RETURN $(\mathcal{G},\ \{(c^x_k,c^y_k)\}_{k=1}^{K})$
\end{algorithmic}
\end{algorithm}

\subsection{Serving Optimization}\label{sec:serving}
Then we perform serving-related optimization comprises (i) \textbf{parallel mapping} and (ii) \textbf{dynamic scheduling} across PEs/chiplets.

\paragraph{Parallel mapping.}
Layers are partitioned into pipeline stages 
for both phases as shown in Fig.~\ref{fig:inference}. We map each stage onto a subset of PEs distributed across chiplets. Within a stage, TP shards a layer across the selected PEs; because inter-PE communication is non-negligible, not every PE should participate in every stage. The assignment space is combinatorial—choosing $TP$ out of $N^{p/d}_{PE}$ PEs is $\binom{N^{\text{p/d}}_{\text{PE}}}{TP}$ per layer—and further multiplies under PD disaggregation.

We propose to cast the mapping problem as a two-stage optimization. \emph{Stage~1 (TP grouping).} For a candidate TP, we solve a small integer linear program (MILP) that clusters all PEs into groups of size $TP$ and nominates a {center PE} per group. The algorithm is shown in Alg.~\ref{alg:ilp_cluster}. The objective is to minimize (i) intra-group all-reduce cost (approximated by the Manhattan diameter and center-based reductions) and (ii) inter-stage transport cost via center-to-center distance. 
\emph{Stage~2 (solve mapping).} Given the TP groups, we assign pipeline stages to groups for both prefill and decode phases using simulated annealing. The objective is to minimize the sum of stage latency and inter-layer KV-transfer latency. 
Figure~\ref{fig:dataflow}(b) illustrates one outcome ((TP,PP)=(8,1) for prefill and (4,2) for decode), consistent with observations in DistServe~\cite{zhong2024distserve} that larger TP benefits prefill and deeper PP improves decode throughput. 
When memory capacity permits, data parallelism (DP) can also be activated to run multiple instances, thus adjusting the PD pool sizes flexibly as SplitWise~\cite{patel2024splitwise}.

\paragraph{Dynamic scheduling.}
At runtime, we adopt iteration-level dynamic scheduling in the spirit of ORCA~\cite{yu2022orca}, with selective batching for both pre-attention (\emph{QKV Proj.}) and post-attention (\emph{O Proj.}, FFNs). As illustrated in Fig.~\ref{fig:dataflow}(c), upon receiving activations (\emph{Act.}), each chiplet multicasts (MC) over the NoC to the first-stage PEs. As soon as \emph{QKV Proj.} begins, the KV cache is forwarded over the NoC–NoP fabric to the paired decode PEs to hide transport stalls. 
The attention block (\emph{Attn.}) and \emph{O Proj.} follow; \emph{O Proj.} triggers an all-reduce decomposed into a \emph{reduce} (RD) and a subsequent \emph{multicast} (MC). The dynamic scheduler overlaps these collectives with downstream compute to sustain utilization under variable request lengths. Together, the mesh-aware parallel mapping and runtime scheduler constitute our serving optimization.

\subsection{System Simulator} \label{sec:simulator}
Our end-to-end simulator (Fig.~\ref{fig:framework}) integrates three levels: \emph{hardware characterization}, \emph{operator modeling}, and \emph{system simulation}. 

\paragraph{Hardware level.}
We build parameterized models for 3D-DRAM, PEs, and interconnects to provide area, power, and performance metrics for design exploration. The 3D-DRAM model considers the bank/layer size, timing constraints, and temperature-dependent refresh; PE models capture MPU and VPU pipelines and basic SRAM behaviors. Interconnect models include NoC and NoP delay/energy.

\paragraph{Operator level.} \label{sec:core-sim}
We quantify the latency and energy of compute, memory, and communication operations under the proposed dataflow. 
For memory, a request of size $DATA$ requires $N_{\text{cmd}} = \frac{DATA}{N_{\text{bank}} \cdot N_{\text{io}} \cdot BL}$
commands per MC, assuming balanced distribution~\cite{poddar2016power}. Each command incurs $t_{\text{mem}} = t_{RCD} + t_{CAS} + t_{RP} + BL \cdot N_{\text{cmd}},$
plus TSV delay under a layer-interleaving scheme for full-bus utilization.  
Refresh overhead is modeled as $t_{\text{refresh}} = \frac{t_{R/W}}{t_{RFI}} \cdot t_{RFC},$
scaling with temperature. Energy is proportional to command count and refresh commands.
For computation, we analytically evaluate cycle counts and MPU utilization following a ScaleSim-style model~\cite{samajdar2019scalesim}; results are stored  in a lookup table (LUT) keyed by tensor shape and tiling strategy. 
For communication, given message size $m$ and hop count $h$, we model delay as $t_{\text{comm}} = \alpha \cdot m + \beta \cdot h$, 
where $\alpha$ and $\beta$ are calibrated link/router parameters~\cite{zhang2024llmcompass,fang2024palm}. Energy scales with $m$ and $h$.

\paragraph{System level.}
We represent batched LLM requests as DAGs of operators. Latency/energy are instantiated from LUT entries, and an event-driven simulator in \texttt{SimPy}~\cite{10.7717/peerj-cs.103} produces fine-grained timestamps for SLO evaluation. The simulator generates PE activity traces; combined with leakage models, these yield power traces that feed a thermal solver. Both transient and steady-state analyses are supported (steady state by default). DRAM static power and refresh are iteratively updated with temperature until convergence. We assume liquid cooling with parameters from~\cite{5361225} and employ temperature-aware flow-rate control, which adjusts coolant flow to trade off pumping power and effective thermal resistance.

\subsection{Design Space Exploration (DSE)} \label{sec:dse}
We design a hierarchical DSE framework (Fig.~\ref{fig:framework}) with two stages: \emph{chiplet-level exploration} and \emph{system-level assembly}.

\paragraph{Chiplet level.}
We sweep DRAM and PE parameters to identify feasible chiplets under area, power, and technology limits. Objectives include peak compute density (TFLOPS), DRAM bandwidth, and capacity. A multi-objective Bayesian Optimization finds Pareto-optimal designs. For each DRAM capacity, Pareto and near-Pareto candidates are selected as inputs to the next stage.

\paragraph{System level.}
We assemble systems from candidate chiplets and jointly optimize the prefill/decode ratio, TP/PP degrees, placement, and runtime scheduling. 
We employ a constrained Bayesian-optimization loop (e.g., expected feasible improvement) that
(i) initializes PD-balanced seed designs and prunes any violating packaging/technology limits (reticle area, package pins, rack/TDP power);
(ii) tunes TP/PP on the two-level mesh (Sec.~\ref{sec:metrics}); in this stage, chiplet placement is also adjusted to obtain a floorplan favorable to parallel mapping;
(iii) evaluates each candidate with the event-driven simulator to obtain \textbf{PPAT} metrics—Performance (TTFT, TBT, throughput), Power (steady and peak), Area, and Thermal (steady), plus capacity;
(iv) filters by SLO and safety constraints (e.g., $\mathrm{TTFT}\!\le\!\mathrm{TTFT}_{\max}$, $\mathrm{TBT}\!\le\!\mathrm{TBT}_{\max}$, $T_{\max}\!\le\!T_{\text{limit}}$, $P_{\text{peak}}\!\le\!P_{\text{rack}}$, capacity $\ge$ KV footprint) and picks new samples to maximize \emph{throughput-per-watt}; and
(v) iterates until convergence, returning the optimal system configuration.

This hierarchical framework links hardware-, architecture-, and system-level choices and yields globally optimized solutions for LLM serving.

\begin{table}[h]
\centering
\vspace{-0.1cm}
\caption{Chiplet design parameters and value sets.}
\footnotesize
\resizebox{0.49\textwidth}{!}{
    \begin{tabular}{cc|cc}
    \hline\hline
    \multicolumn{4}{c}{Distinctive Params. for PD: \#$10^5$} \\ \hline
    \textbf{Parameters} & \textbf{Vals} & \textbf{Parameters} & \textbf{Vals} \\ \hline
    DRAM IO Width & [32, 64, 128, 256, 512] 
    & DRAM Capacity/GB & [1, 2, 4, 8, 16, 32] \\ \hline
    DRAM $N_{\text{layer}}$ & [1, 2, 3, 4, 5] 
    & DRAM banks & [8, 16, 32, 64, 128] \\ \hline
    Page Size & [1024, 2048, 4096, 8192]
    & Cores & [1,2,4,8,10,16,24,32] \\ \hline
    PEs  & \multicolumn{3}{c}{[4, 6, 8, 9, 10, 12, 16, 18, 20, 24, 25]} \\ \hline
    \multicolumn{4}{c}{Common Params: \#$10^5$} \\ \hline
    \textbf{Parameters} & \textbf{Vals} & \textbf{Parameters} & \textbf{Vals} \\ \hline
    SRAM banks & [4, 8, 16, 32] 
    & SRAM capacities/Kb & [64,128,256,512,1024,2048]\\ \hline
    SArows & [16, 32, 64, 128] 
    & baseSArows & [1, 2, 4, 8, 16] \\ \hline
    SAcols & [16, 32, 64, 128] 
    & Vector registers & [16, 32, 64, 128] \\ \hline
    NoC flits & [128, 256, 512, 1024, 2048] 
    & NoP channels & [2,4,6,8,10,12] \\ \hline\hline
     \label{tab:params}
    \end{tabular}
}
\vspace{-0.3cm}
\end{table}

\section{Evaluation}\label{sec:evaluation}

\subsection{Experimental Setup}\label{sec:exp-setup}

\paragraph{Benchmarks.}
We evaluate three dense models—GPT-13B, QwQ-32B, and LLaMA3-70B—adopting MHA or GQA as appropriate.
Workloads come from three datasets: (i) \textit{Code} (mean input/output lengths: 2071/25), production traces from Azure~\cite{patel2024splitwise}; (ii) \textit{Reason} (1473/1293), synthesized traces based on the DeepSeek-R1 report~\cite{zhihu2025}; and (iii) \textit{LongBench} (7108/5), targeting long-context tasks~\cite{bai2023longbench}.
Each dataset uses a 200-request trace and captures a distinct task regime.

\paragraph{System configurations.}
\emph{LaMoSys3.5D} logic dies use a 7\,nm process and run at 800\,MHz. Key design parameters appear in Tab.~\ref{tab:params}. HB-related area figures are derived from our tape-out chips~\cite{niu2022184qps} and synthesized for 128/256/512/1024 pins.
A representative system instantiates $5\times$ \emph{PCs} and $4\times$ \emph{DCs} with the floorplan in Fig.~\ref{fig:template}.
Each \emph{PC} integrates 16 PEs (16 cores/PE), 32 GB of four-stacked DRAM delivering 10.6 TB/s, and sustains 400 TFLOPS peak compute.
Each \emph{DC} integrates 32 PEs (8 cores/PE), 64 GB of eight-stacked DRAM at 26.2 TB/s, and the same peak compute.
Die areas are $\SI{546}{mm^2}$ (PC) and $\SI{584}{mm^2}$ (DC); peak powers are 438 W and 638 W, yielding compute densities of $\sim1$ TFLOPS/W. DRAM energy is $\sim0.7$ pJ/bit.
The intra-chiplet NoC and inter-chiplet die-to-die bandwidths are 200 GB/s and 800 GB/s, respectively.

\paragraph{Baselines.}
We compare \emph{LaMoSys3.5D} against four serving platforms:
\textbf{A100} (six nodes: two prefill, four decode; each node has $4\times$ A100 with 80 GB HBM3, connected by NVLink),
\textbf{TPU} (eight nodes: four prefill, four decode; each node has $4\times$ TPUv4 with 32 GB HBM3),
\textbf{LC-L} (eight nodes: four prefill, four decode; $4\times$ LC-L per node), and
\textbf{LC-T} (eight nodes: four prefill, four decode; $2\times$ LC-T per node).
We also compare three 3D-DRAM architectures—(i) TETRIS~\cite{gao2017tetris}, (ii) 3D-TokSIM~\cite{20253dtoksim}, and (iii) 3D-LC~\cite{sharda2024accelerator}—using LLMCompass-type dataflows. When certain parameters are missing from the paper, we perform targeted calibration for fair comparison.
All designs are normalized to a 7\,nm logic process for consistency.

\paragraph{Simulation.}
We adapt CACTI-3DD~\cite{6176428} to model 3D-DRAM timing/physical parameters and calibrate against tape-out silicon. 
Logic components are modeled using a customized NeuroMeter~\cite{9407039} to obtain power/area for MPUs, VPUs, SRAM, and routers; metrics for other blocks (e.g., MC and RV controller) follow~\cite{chen2024high}.
The NoC uses McPAT~\cite{5375438} to model the router and to derive link delays and bandwidth. The leakage model follows the calibrated McPAT~\cite{TCAD2023-McPAT-Calib}.
For the NoP, we refer to HISIM studies~\cite{10844846,chipsalliance-aib}. 
Thermal behavior is simulated with modified ATSim~\cite{wang2024atsim3d}, with parameters from in-house chips.
Because most public results are architecture-level, we develop an in-house simulator following DistServe~\cite{zhong2024distserve} and GenZ~\cite{xia2024agentless} to evaluate serving performance across all systems, and we search TP/PP settings to obtain the best configuration for each platform.

\begin{figure*}
    \centering
    \includegraphics[width=1.03\linewidth]{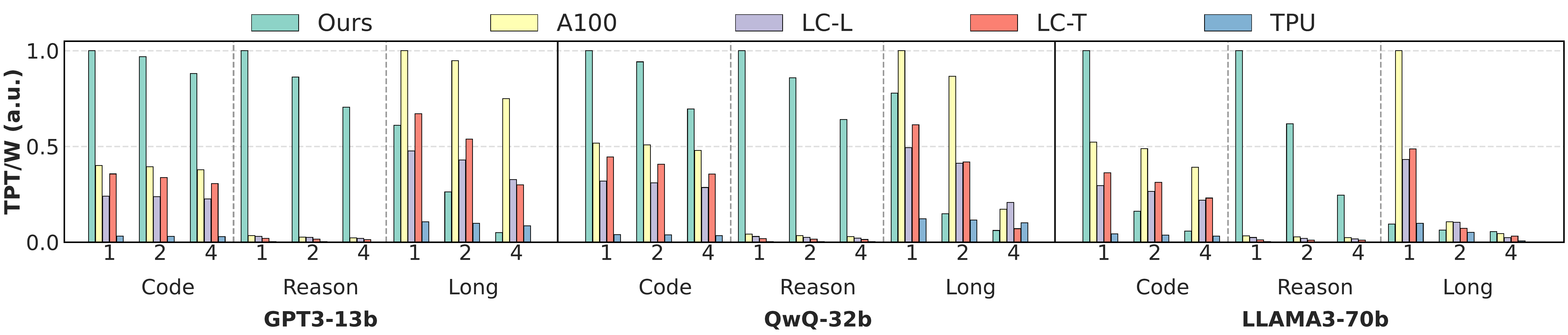}
    \caption{Comparison of inference performance across systems. Labels `1,2,4' in the X axis represent the request rate.}
    \label{fig:inference-models-hardware}
\end{figure*}

\begin{figure}
    \centering
    \includegraphics[width=.94\linewidth]{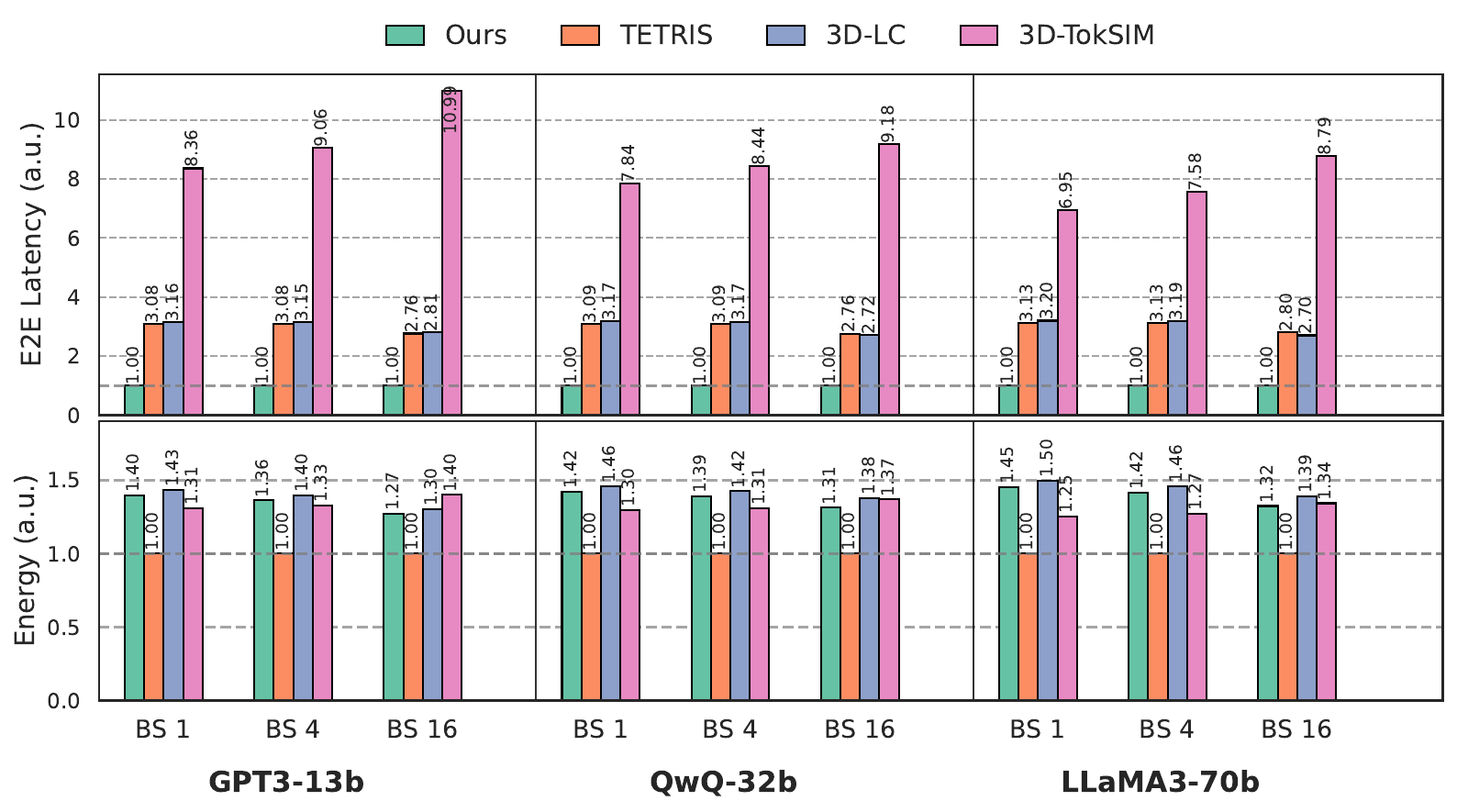}
    \caption{Comparison across 3D Architectures.}
    \label{fig:inference-3d-hardware}
\end{figure}
\vspace{-0.2cm}
\begin{figure}[tbh]
    \centering
    \includegraphics[width=.94\linewidth]{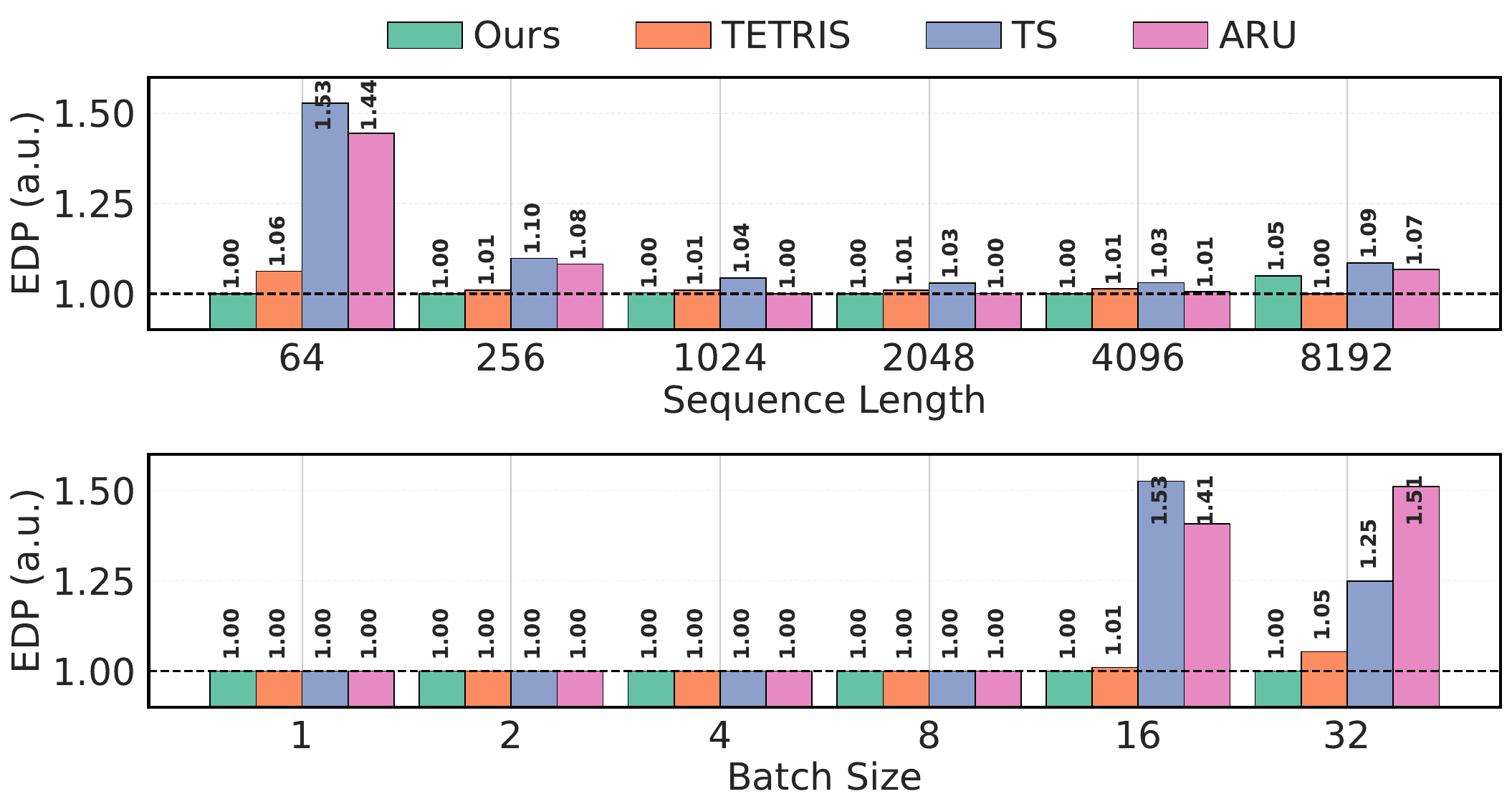}
    \caption{Comparison of dataflow design.}
    \label{fig:dataflow-compare}
\end{figure}
\begin{figure}[tbh]
    \centering
    \includegraphics[width=\linewidth]{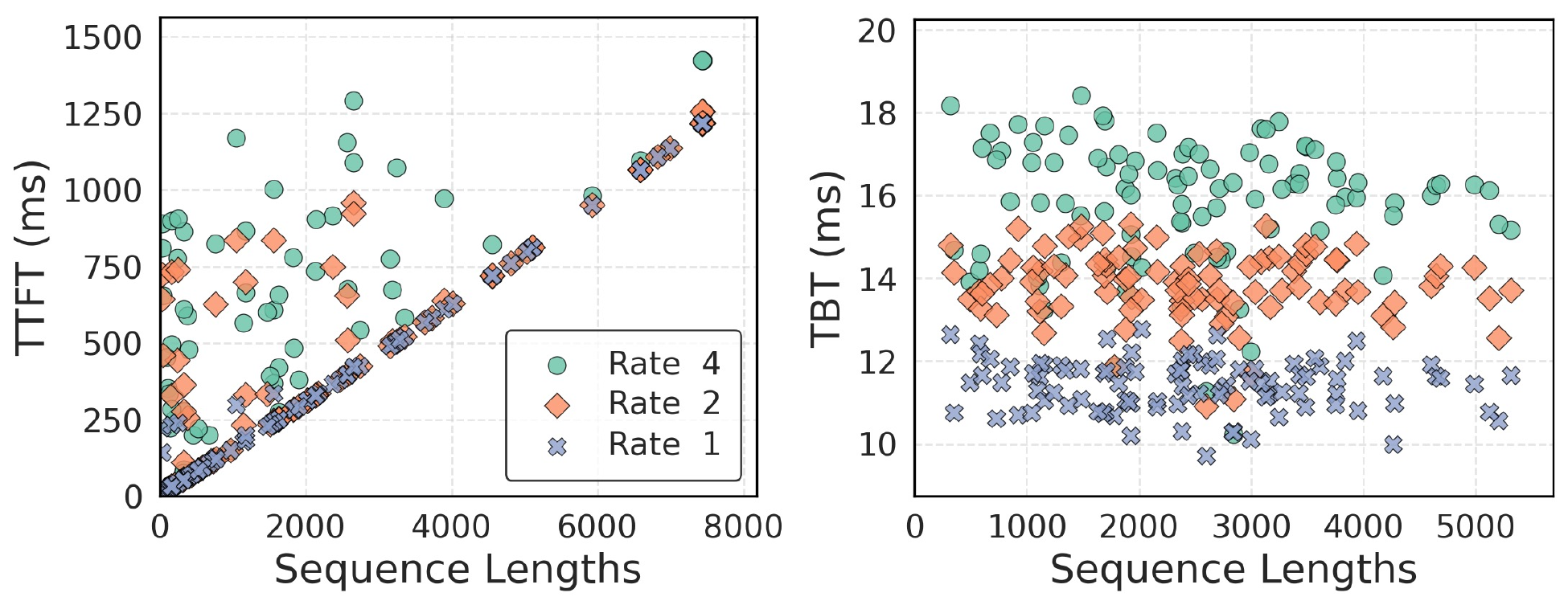}
    \caption{Serving performance under different request rates.}
    \label{fig:rate}
\vspace{-0.4cm}
\end{figure}

\subsection{Baseline Comparison}\label{sec:baseline-comp}
\paragraph{Comparison with Inference Systems}
We evaluate the optimal \emph{LaMoSys3.5D} architecture against baselines at request rates of 1, 2, and 4~req/s. 
Fig.~\ref{fig:inference-models-hardware} reports \emph{throughput per watt} (tokens/s/W), which jointly reflects performance and energy efficiency. 
Because runtime power for each system is unavailable, we use the published maximum design power for all systems—including ours—and sum over participating chips.
\emph{LaMoSys3.5D} reaches \textbf{0.75}~tokens/s/W, exceeding A100 (\textbf{0.46}), TPUv4 (\textbf{0.05}), LC-L (\textbf{0.24}), and LC-T (\textbf{0.29}). 
Two factors drive this gain: 
(i) lower total power, primarily from reduced memory energy and smaller SRAM footprint, while compute efficiency remains comparable to A100 (about 1–1.2~TFLOPS/W); 
and (ii) higher throughput—\emph{LaMoSys3.5D} averages \textbf{3{,}466}~tokens/s versus \textbf{3{,}370} for A100 and \textbf{1{,}023} for TPUv4.
\emph{LaMoSys3.5D} also delivers the lowest \textbf{TBT} across workloads, improving by \textbf{17$\times$} (geomean) over DGX-A100 and \textbf{73$\times$} over TPUv4. 
This stems from much higher per-device memory bandwidth (e.g., \emph{DC} chiplet at 26.2~TB/s vs.\ A100 at $\sim$2~TB/s) and the $D^3$ dataflow that exploits 3D-DRAM bandwidth. 
By contrast, the TPUv4 is simulated following LLMCompass’s default dataflow, which may be less efficient for GEMV-dominant decode on large systolic arrays.
Prefill is modestly slower: average \textbf{TTFT} is $\sim$2.13$\times$ A100 due to fewer compute units. 
For reference, LC-L, LC-T, and TPUv4 exhibit $\sim$1.63$\times$, $\sim$2.33$\times$, and $\sim$3.63$\times$ A100’s \textbf{TTFT}, respectively. 
Except for \textit{LongBench} cases with input length $>$10{,}000, \textbf{TTFT} stays within typical SLOs (sub-second to a few seconds). 
For the \textit{Reason} workload with long generations ($\sim$1{,}300 tokens), \textbf{TBT} gains are even larger. 

\paragraph{Comparison with 3D Architectures}
We further compare \emph{LaMoSys3.5D} with existing 3D-DRAM designs (TETRIS, 3D-LC, 3D-TokSIM). Since these target single-transformer acceleration, we evaluate mean end-to-end latency and energy across multiple models, with input/output lengths ranging from 64 to 8192, and batch sizes \{1,4,16\} (Fig.~\ref{fig:inference-3d-hardware}).
\textbf{Latency.} \emph{LaMoSys3.5D} achieves the lowest E2E latency. Averaged over all settings and normalized to \emph{LaMoSys3.5D}, the mean latencies are $2.99\times$ (TETRIS), $3.02\times$ (3D-LC), and $8.58\times$ (3D-TokSIM), thanks to the $D^3$ dataflow and its search over the full mapping space.
\textbf{Energy.} TETRIS attains the lowest mean energy due to its DRAM-access–minimizing dataflow. Normalized to TETRIS, the mean energies are $1.37\times$ (\emph{LaMoSys3.5D}), $1.32\times$ (3D-TokSIM), and $1.41\times$ (3D-LC). Memory dominates energy across designs, while compute energy is similar.

To conclude, \emph{LaMoSys3.5D} delivers superior system-level throughput and energy efficiency versus inference baselines, and the best single-request latency among 3D designs—driven by high 3D-DRAM bandwidth and effective utilization via our dataflow and parallel mapping design. The results also indicate that thermal headroom is a first-order design constraint in 3.5D-ICs, necessitating the incorporation of power and thermal budgets across diverse workloads.

\subsection{Performance Analysis}\label{sec:dataflow-thermal}
\paragraph{Dataflow.}
Fig.~\ref{fig:dataflow-compare} compares energy--delay product (EDP) across four dataflows:
(i) the proposed \textbf{$D^3$};
(ii) \textbf{TETRIS}~\cite{gao2017tetris};
(iii) \textbf{TS}: token-stationary~\cite{20253dtoksim};
(iv) \textbf{ARU}: SRAM-reuse--centric.
\textbf{Prefill} $D^3$ achieves the lowest EDP; averages relative to $D^3$ are $1.01\times$ (TETRIS), $1.13\times$ (TS), and $1.09\times$ (ARU). As sequence length grows, the EDP for all DRAM-involved flows converges because computation dominates.
\textbf{Decode} EDP of all flows is comparable at small batch sizes. For larger batches (e.g., 16/32), TS and ARU degrade in both latency and energy due to limited SRAM capacity, which increases 3D-DRAM traffic. $D^3$ sustains low EDP by allocating reuse across 3D-DRAM and SRAM flexibly, thus maintaining high vertical bandwidth utilization.
Across regimes, $D^3$ delivers the best EDP and scales more stably with batch size and sequence length. These results underscore that a single parameterized hardware template can instantiate compute-rich prefill chiplets and bandwidth-/capacity-rich decode chiplets, supporting heterogeneous specialization without sacrificing design coherence.

\paragraph{Serving.}
Figure~\ref{fig:rate} shows latency distributions under varying request rates. When rates are low, TTFT matches the ideal values, indicating high utilization ($\sim$80\%). At higher rates, shorter requests suffer queuing delays behind longer ones, inflating TTFT. To mitigate this issue, techniques such as preemption and chunked prefill \cite{agrawal2024taming} can be adopted in the future.
For decoding, TBT scales with batch size; longer outputs or higher request rates enlarge decoding batches, thus raising TBT. To optimize throughput while adhering to SLOs, it is advisable to enlarge the batch size without compromising TBT. 

\begin{figure*}[tbh]
    \centering
    \subfloat[Code]{
        \includegraphics[width=0.3\linewidth]{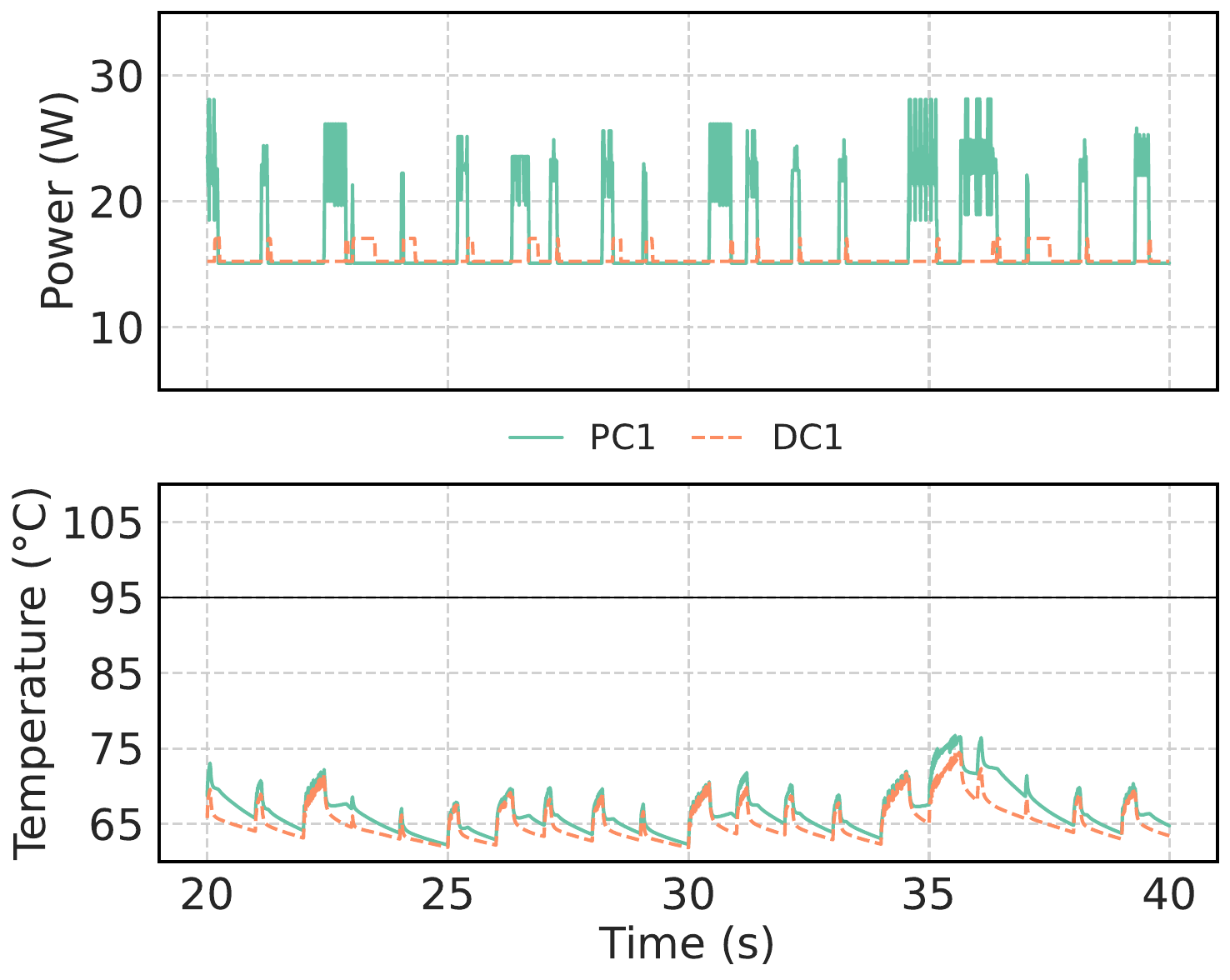}
    }\hfill
    \subfloat[Reason]{
        \includegraphics[width=0.3\linewidth]{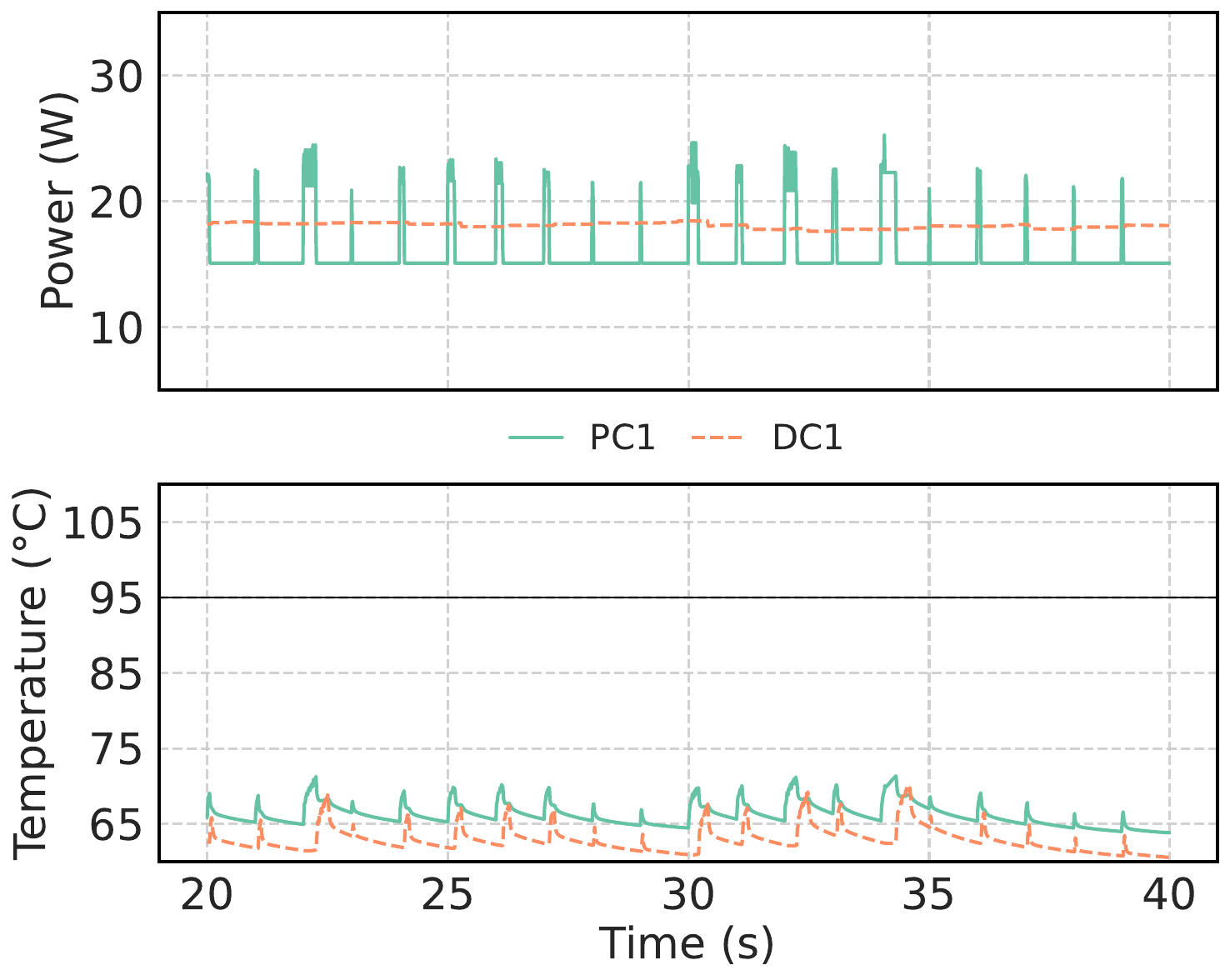}
    }\hfill
    \subfloat[Longbench]{
        \includegraphics[width=0.3\linewidth]{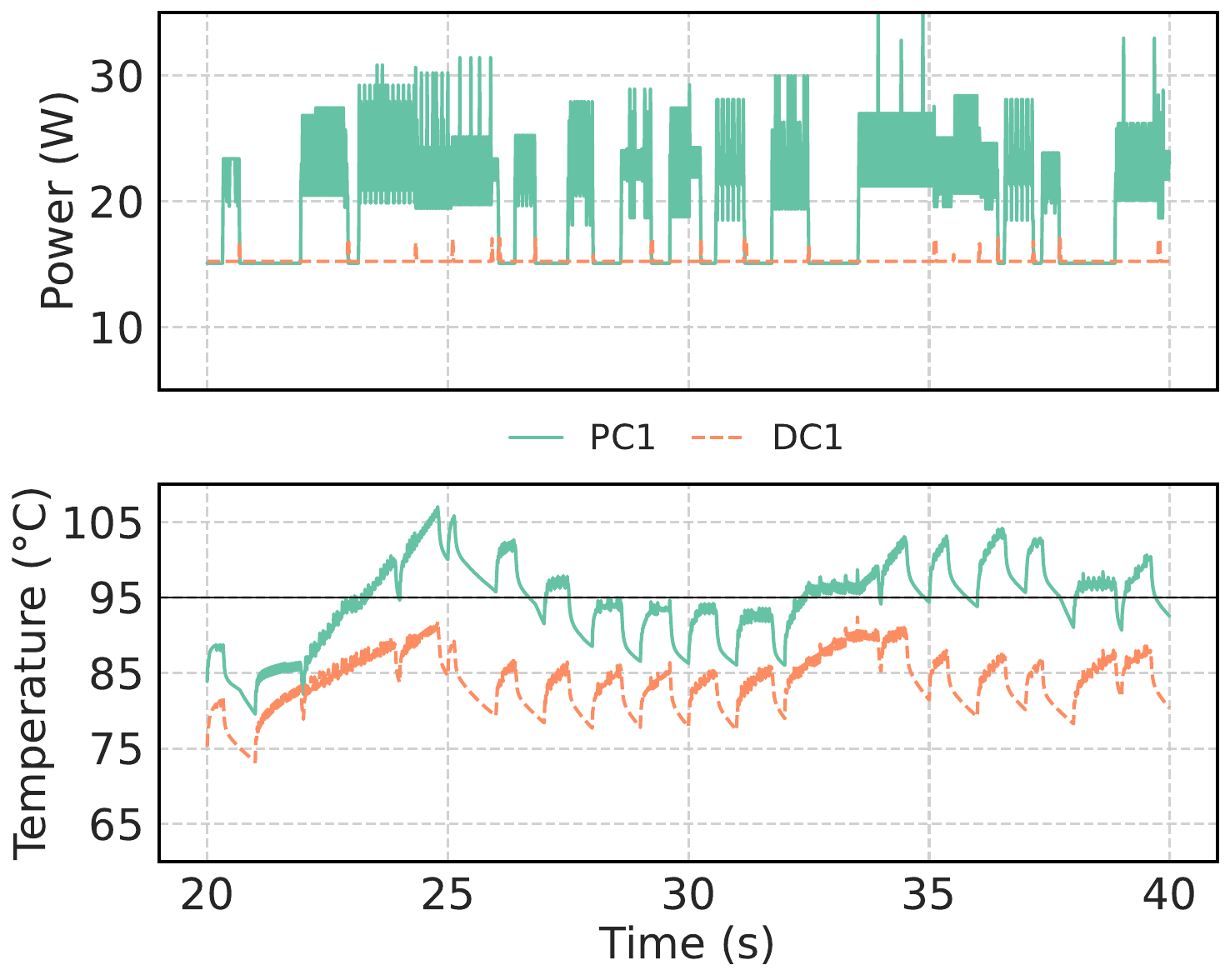}
    }\hfill
    \caption{Power and temperature traces for the LaMoSys3.5D running workload `Code'(a),  `Reason'(b),  `Long'(c).}
    \label{fig:power-thermal-trace}
    \vspace{-0.2cm}
\end{figure*}

\paragraph{Thermals and Power.}
We select one PE from a prefill chiplet (PC1) and one from a decode chiplet (DC1) and trace their power and center-point temperature over time under multiple workloads (Fig.~\ref{fig:power-thermal-trace}). All results assume liquid cooling with an ambient temperature of \SI{45}{\celsius}. 
For the PC, irregular input lengths create idle gaps and stalls; power spikes during active intervals and drops when idle, producing rapid temperature rises followed by gradual decay between bursts. 
In contrast, the DC operates nearly continuously with lower compute intensity; power varies less and temperature is more stable. Despite its lower power, the DC can match or exceed the PC’s temperature because it integrates eight DRAM layers (vs.\ four on the PC), lengthening the heat path and increasing effective thermal resistance. 
Under \textit{Reason} and \textit{Code}, temperatures settle near $\sim$\SI{70}{\celsius}; with \textit{LongBench}, stronger prefill demand pushes temperatures toward $\sim$\SI{100}{\celsius}. These results underscore the need for workload-aware, transient thermal evaluation early in architecture design (Sec~\ref{sec:motiv-thermal}): power and thermal profiles vary sharply across workloads, and nominal or worst-case static assumptions can misestimate bandwidth, latency, and energy.

\subsection{DSE Analysis}\label{sec:dse-results}
In this subsection, we explore the architecture design space of \emph{LaMoSys3.5D} and summarize several key architectural implications for future design. Since all evaluated models show consistent trends across architectural parameters, we present results for QwQ-32B as a representative case without loss of generality.

\begin{figure}[tbh]
    \centering
    \subfloat[Chiplet-DSE]{
        \includegraphics[width=0.46\linewidth]{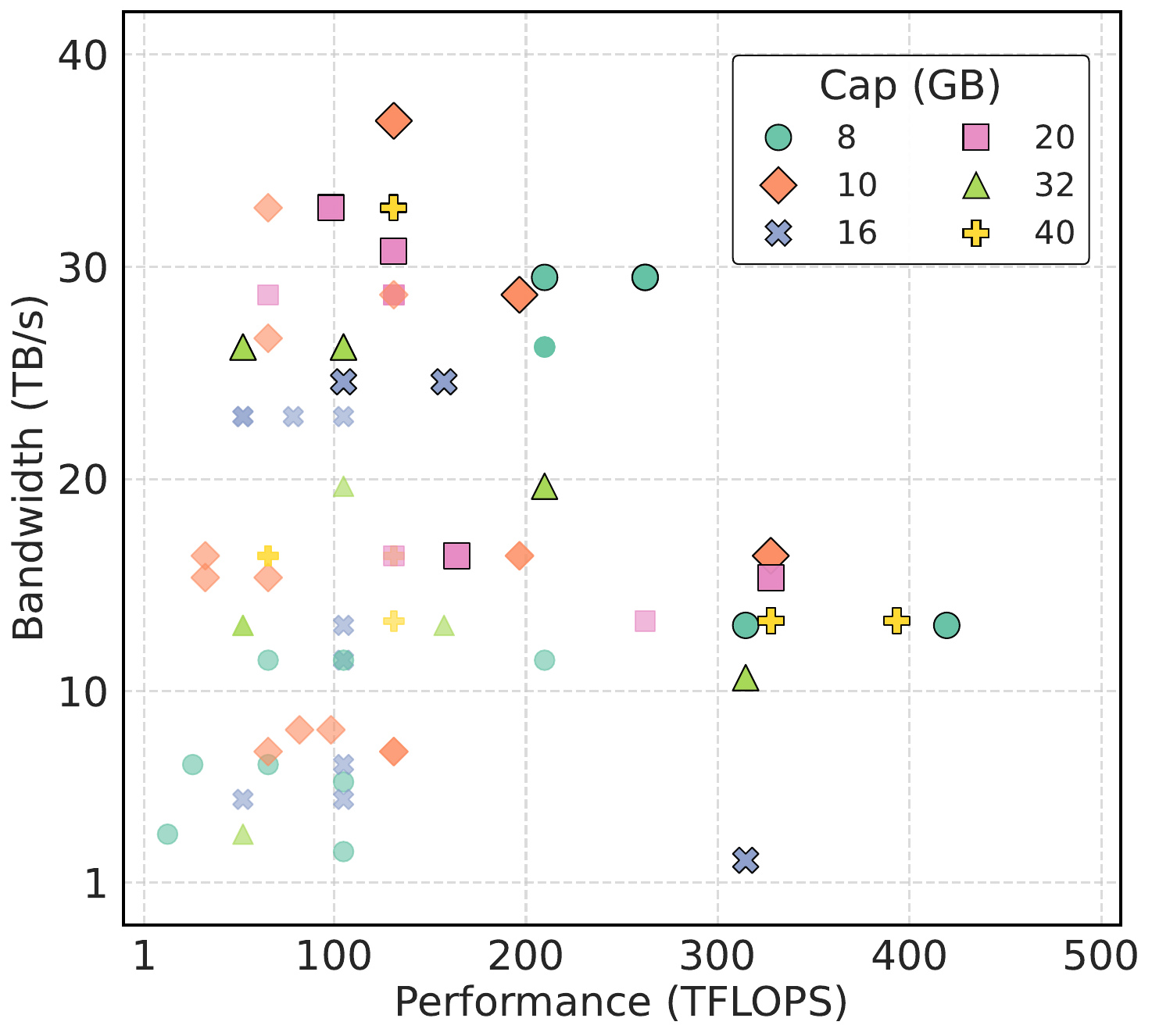}
    }\hfill
    \subfloat[System-DSE]{
        \includegraphics[width=0.47\linewidth]{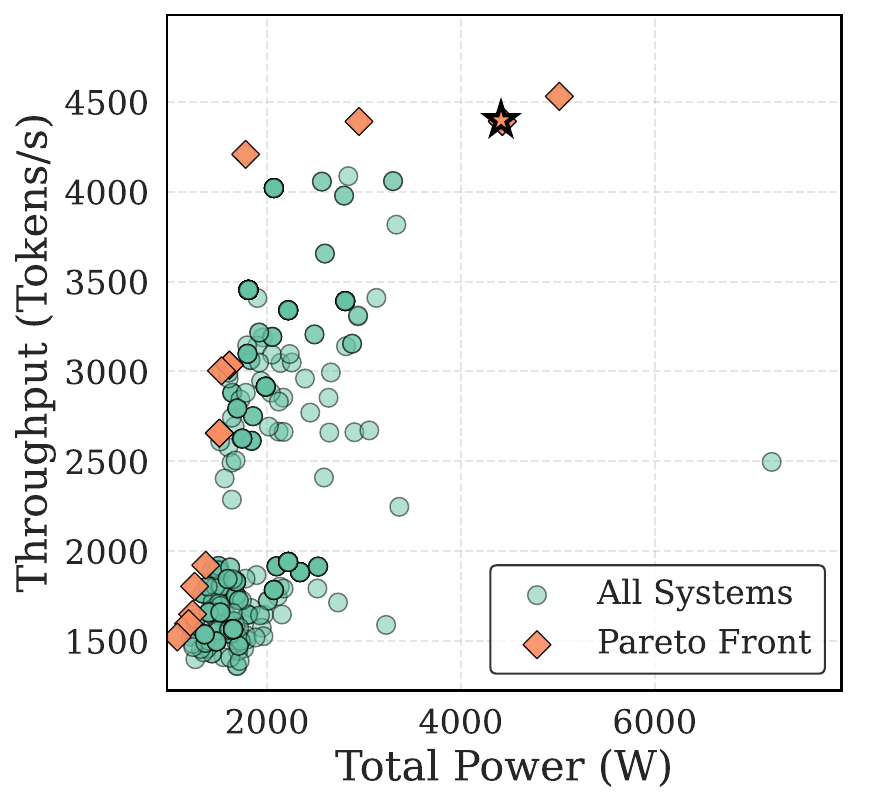}
    }
    \caption{DSE results for chiplet (a) and system (b).}
    \label{fig:dse}
    \vspace{-0.2cm}
\end{figure}

\paragraph{System design trade-off.}
Figure~\ref{fig:dse}(a) shows the performance--bandwidth distribution of chiplets with different capacities, revealing a clear trade-off between these two aspects. This observation validates the rationale in Sec.~\ref{sec:motivation}: allocating higher bandwidth to decoding while dedicating more compute resources to prefill. 
After identifying feasible chiplet designs, we proceed to system-level optimization. Results are shown in Figure~\ref{fig:dse}(b). We present throughput and power distributions, where Pareto-optimal designs are highlighted. Notably, power is included as an optimization target, explaining why some designs with large TTFT and TBT remain on the Pareto front. We select the design marked by \ding{80} as the final design choice for its larger memory capacity.

\begin{figure}[tbh]
    \centering
    \includegraphics[width=.98\linewidth]{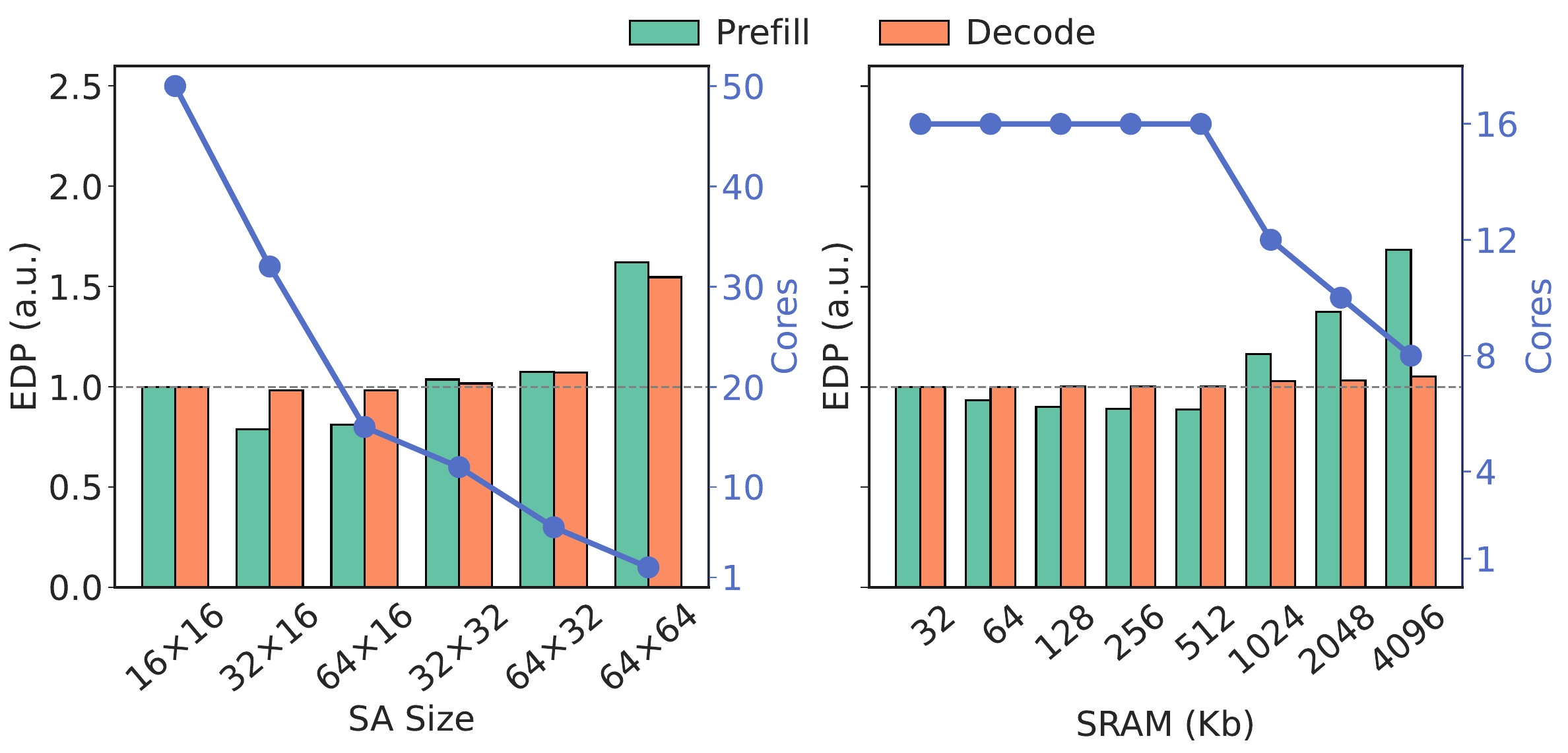}
    \caption{Impact of MPU size and SRAM buffer capacity on prefill and decoding performance.}
    \label{fig:core-sweep}
    \vspace{-0.2cm}
\end{figure}

\paragraph{Core design.}
Fig.~\ref{fig:core-sweep}(a) varies the systolic-array (SA) shape while fixing the number of baseSAs at~8 and holding PE area constant (core count scales accordingly). Smaller \emph{columns} are preferred because they shrink the SA area and permit more cores per PE. Larger \emph{rows} improve prefill by raising per-core throughput. Increasing the number of base SAs further boosts performance—most notably for \emph{Logit}. \emph{Attend} gains less when the head dimension $N$ is small (e.g., 128) and the sequence length $K$ is large, since partial-sum accumulation dominates. \emph{Projection} speedups scale primarily with batch size.
Fig.~\ref{fig:core-sweep}(b) sweeps per-core SRAM while adjusting core count to keep PE size fixed. Medium SRAM sizes are preferred in both phases: $D^3$ has limited reliance on large buffer size, and smaller SRAM size reduces core area, enabling more cores per PE. Overall, balancing per-core capability and core count is crucial—over-sized or under-sized compute arrays and SRAM sizes both hurt performance; a balanced SA shape and medium SRAM per core work best.

\begin{figure}[tbh]
    \centering
    \includegraphics[width=1\linewidth]{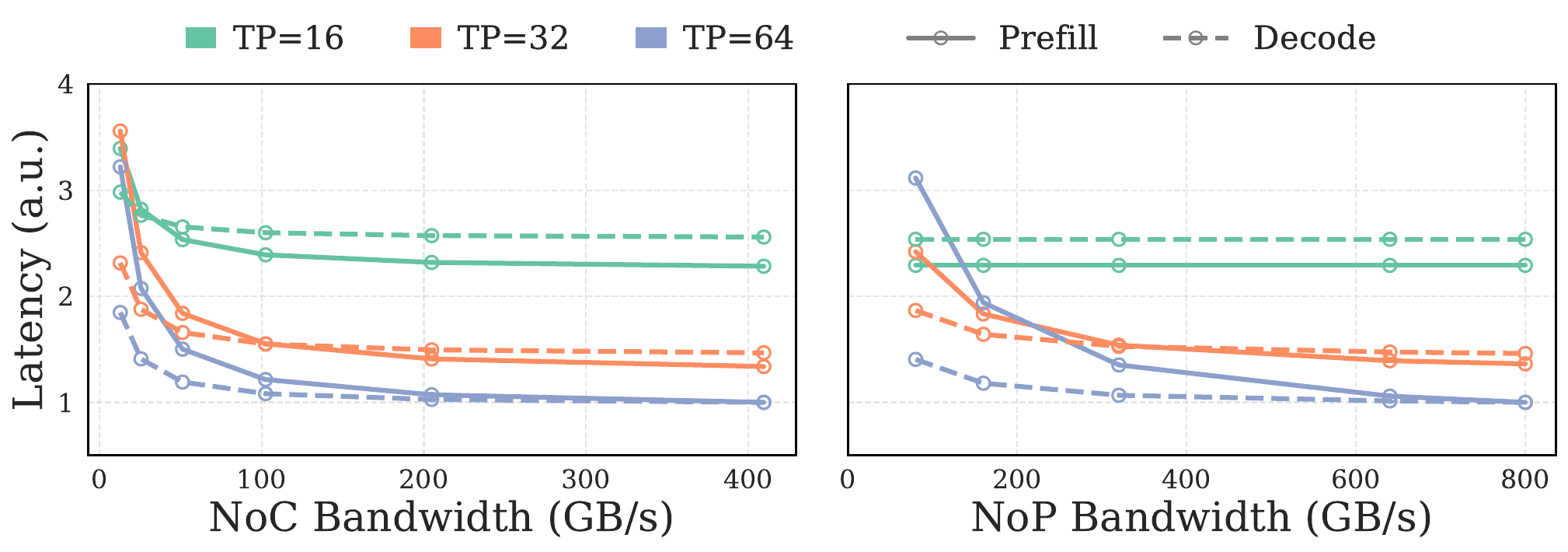}
    \caption{Impact of NoC/NoP bandwidth and TP strategy on prefill and decoding performance.}
    \label{fig:tp-bw}
    \vspace{-0.2cm}
\end{figure}

\begin{figure}[tbh]
    \centering
    \includegraphics[width=.94\linewidth]{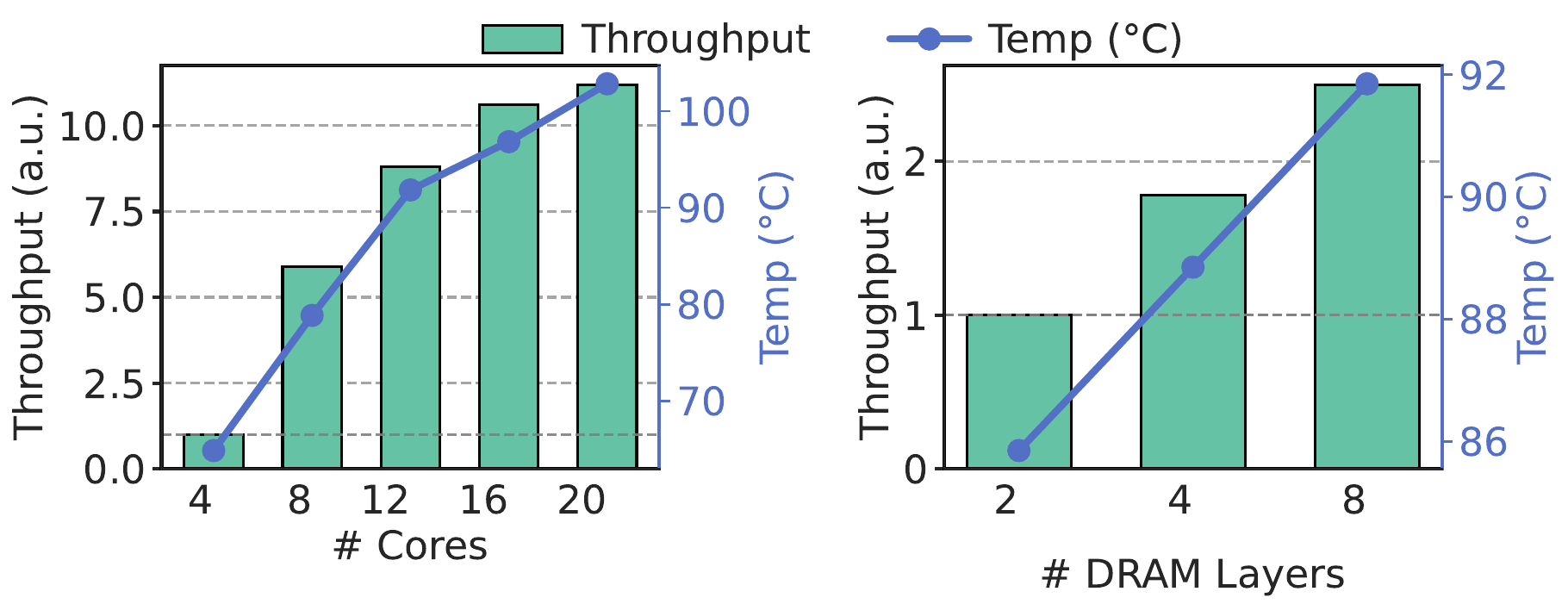}
    \caption{Impact of number of cores and DRAM layers on throughput and temperature.}
    \label{fig:pe-temp}
\end{figure}

\paragraph{Communication design.}
We further examine the impact of NoC/NoP bandwidth and parallel strategies on TTFT and TBT. Figure~\ref{fig:tp-bw}(a),(b) present normalized latency results for NoC and NoP, respectively. With sufficient bandwidth, a larger TP is preferred since it reduces per-PE computation. However, under limited bandwidth, a smaller TP is favored due to the high communication overhead of all-reduce. 
These results highlight the importance of co-optimizing TP and PP (Sec.~\ref{sec:dataflow}). Larger TP reduces compute but increases collective costs; with constrained NoC bandwidth, moderate TP minimizes TTFT. Increasing NoC/NoP bandwidth improves both TTFT and TBT, with TTFT being more sensitive to bandwidth, while a smaller TP is preferred under bandwidth constraints.

\paragraph{PE design.}
Finally, we vary the number of cores per PE (for PC) and the number of DRAM layers (for DC). Fig.~\ref{fig:pe-temp}(a) and (b) report the throughput and maximum temperature. 
Throughput increases as cores/PE grow, but the gain tapers once cores $>16$, when $T_{\max}$ exceeds \SI{95}{\celsius}; the elevated temperature raises DRAM refresh activity, reducing effective memory bandwidth and throttling performance. Increasing the number of layers consistently improves throughput, and the associated temperature rise is modest. Thus, we recommend preferring more DRAM layers while capping the core count (e.g., $\leq 16$), as throughput is highly sensitive to tight thermal budgets.

\section{Related Work}\label{sec:relatedwork}

\subsection{Efficient LLM Inference Serving}
LLM inference is optimized at both engine and service layers~\cite{zhou2024survey,kim2023full}. Engine-level techniques include graph optimization~\cite{dao2023flashattention2}, operator fusion~\cite{dao2022flashattention,hong2023flashdecoding}, offloading~\cite{sheng2023flexgen}, and fast decoding (e.g., speculative decoding)~\cite{leviathan2023fast,cai2024medusa}. Service systems focus on memory management~\cite{kwon2023efficient}, batching~\cite{yu2022orca,agrawal2023sarathi}, scheduling~\cite{yu2022orca,agrawal2024taming}, and distributed execution~\cite{patel2024splitwise,zhong2024distserve}; notable systems include vLLM~\cite{kwon2023efficient} and SGLang~\cite{zheng2024sglangefficientexecutionstructured}. Architectural gains also come from model choices (e.g., GQA~\cite{ainslie2023gqa}) and compression (quantization~\cite{frantar2022gptq}, sparsity~\cite{ma2023llm}). These advances complement our 3.5D-IC design.

\subsection{PIM- and NMP-based Transformer Accelerators}
PIM/NMP architectures exploit internal memory bandwidth and reduce data movement, benefiting LLMs. HBM-PIM targets KV-cache processing in batched inference~\cite{choi2023unleashing}. Recent designs combine 3D-DRAM with PIMs~\cite{zhou2022transpim,heo2024neupims,ding2023haima,seo2024ianus,park2024attacc,2024H3DTransformer,20253dtoksim,2024nicepim}, while 3D-DRAM NMP brings compute near memory~\cite{sharma2023heterogeneous,kim2024monde,sharda2024accelerator}. Some target new regimes, e.g., speculative~\cite{li2024specpim} and long-context~\cite{kwon2024lol}. Most, however, optimize kernels or edge scenarios rather than end-to-end serving. 
\section{Conclusion}\label{sec:conclusion}
With the growing popularity of inference-time computing, there is an increasing demand for more powerful inference systems. In this work, we present our solution: \emph{LaMoSys3.5D}, the \textbf{ first 3.5D-IC architecture designed for efficient LLM serving}. We introduce a novel hardware design to optimize both prefill and decode phases and develop a software framework incorporating features such as dataflow and parallel mapping, thermal-aware modeling, and optimization.
Across diverse LLMs and workloads, \emph{LaMoSys3.5D} improves throughput-per-watt over DGX-A100 systems by 62\% and achieves a $4.87\times$ better end-to-end latency (geo-mean) versus prior 3D designs. It also demonstrates high performance on workloads with long output sequences, 17.0$\times$ decode acceleration compared to A100, making it a promising candidate for future inference systems.

\bibliographystyle{IEEEtran}
\bibliography{./ref/background,./ref/hardware}

\end{document}